\documentclass[aps, longbibliography,twocolumn,showpacs,amsmath,amssymb,superscriptaddress,pra]{revtex4-1}
\usepackage[utf8]{inputenc}

\usepackage{tikz}
\usepackage{graphicx}
\usepackage{amsfonts}
\usepackage{amsmath}
\usepackage{amssymb,bbold}
\usepackage{color}
\usepackage{bm}
\usepackage{bbm}
\usepackage{dsfont}
\usepackage{enumerate}
\usepackage[normalem]{ulem}
\usepackage{amsfonts, amsmath, amsthm, amssymb} 
\usepackage{mathtools}
\usepackage{array}
\usepackage[colorlinks,bookmarks=false,citecolor=blue,linkcolor=red,urlcolor=blue]{hyperref}

\begin{document}

\title{Fractal ground state of {mesoscopic} ion chains in periodic potentials}

\author{Rapha\"el Menu}
\affiliation{Theoretische  Physik,  Universit\"at  des  Saarlandes,  D-66123  Saarbr\"ucken,  Germany}
\author{Jorge Yago Malo}
\affiliation{Dipartimento di Fisica Enrico Fermi, Universita di Pisa and INFN, Largo B. Pontecorvo 3,
I-56127 Pisa, Italy.}
\author{Vladan Vuleti\'c}
\affiliation{Department of Physics, MIT-Harvard Center for Ultracold Atoms, and Research Laboratory of Electronics, Massachusetts Institute of Technology, Cambridge,
Massachusetts 02139, USA.}
\author{Maria Luisa Chiofalo}
\affiliation{Dipartimento di Fisica Enrico Fermi, Universita di Pisa and INFN, Largo B. Pontecorvo 3,
I-56127 Pisa, Italy.}
\author{Giovanna Morigi}
\affiliation{Theoretische  Physik,  Universit\"at  des  Saarlandes,  D-66123  Saarbr\"ucken,  Germany}

\date{\today}

\begin{abstract}
Trapped ions in a periodic potential are a paradigm of a frustrated Wigner crystal. The dynamics is captured by a long-range Frenkel-Kontorova model. We show that the classical ground state can be mapped to the one of a 
{ long-range Ising} spin chain 
in a magnetic field, whose strength is determined by the mismatch between chain's and substrate lattice's periodicity. The mapping is exact when the substrate potential is a piecewise harmonic potential and holds for any two-body interaction decaying as $1/r^\alpha$ with the distance $r$. The ground state is a devil's staircase of regular, periodic structures as a function of the mismatch  and of the interaction exponent $\alpha$. While the staircase is well defined in the thermodynamic limit for $\alpha>1$, for Coulomb interactions, $\alpha=1$, { we argue} that it disappears and the sliding-to-pinned transition becomes a crossover{, with a convergence to the thermodynamic limit scaling logarithmically with the chain's size. Due to this slow convergence,  fractal properties can be observed even in chains of hundreds of ions at laser cooling temperatures.} 
These dynamics are a showcase of the versatility of trapped ion platforms for exploring the interplay between frustration and interactions.
\end{abstract}

\maketitle

\section{Introduction}
Chains of laser-cooled ions in linear Paul traps are paradigmatic realizations of a harmonic crystal in one dimension \cite{Dubin:RMP}. In these systems, order emerges from the interplay between the Coulomb repulsion and the trapping potential. Even in one dimension, the long-range nature of Coulomb interactions warrants diagonal (quasi) long-range order, and any finite chain is effectively a one-dimensional Wigner crystal \cite{Schulz:1993}. At the typical temperatures reached by laser cooling the ions {vibrate} harmonically at the crystal equilibrium position and their motion is described by an elastic crystal with power-law coupling \cite{Morigi:2004}. The experimental capability to image and monitor the individual ions makes ion chains a prominent platform for studying structural phase transitions \cite{Birkl:1992,Raizen:1992,Dubin:1993,Fishman:2008} and the static and dynamic properties of crystal dislocations \cite{Ulm,Pyka,Mielenz,Ejtemaee:2013,Brox:2017,Kiethe:2017,Kiethe:2018,Gangloff:2022}. {The} progress in cooling and trapping \cite{Eschner:2003} paves the way for investigating these dynamics deep in the quantum regime \cite{Shimshoni:2011,Zhang:2023,Bonetti,Timm:2021,Vanossi:2013,TosattiPNAS}. 

\begin{figure}[b]
    \centering
    \includegraphics[width=\columnwidth]{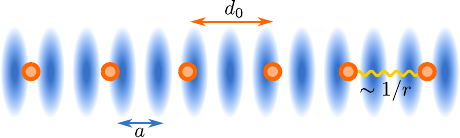}
    \caption{Illustration of a trapped-ion realization of the Frenkel-Kontorova model. 
    {In the absence of other external potentials, c}old ions (orange circles) {confined on a line form a chain} at uniform distance $d_0$ minimizing the Coulomb repulsion. {Geometric frustration is introduced by
    the} standing wave of a laser field (blue shades), forming a periodic potential with periodicity $a$. For $d_0\gg a$ the ions' motion about the equilibrium positions is described by their harmonic vibrations about the equilibrium configuration. The dynamics is captured by a Frenkel-Kontorova model with long-range elastic forces.}
    \label{fig:1}
\end{figure}

Interfacing ion chains with optical lattices, as illustrated in Fig.\ \ref{fig:1}, implements a simulator of nanofriction \cite{Vanossi:2013}. In fact, the Hamiltonian can be reduced to an extended Frenkel-Kontorova (FK) model \cite{Garcia-Mata,Pruttivarasin:2011,Cetina:2013,Cormick:2013}. The FK model describes the interaction of an elastic crystal with an underlying periodic substrate in one dimension \cite{Braun_Kishvar}. Frustration emerges from the mismatch between the periodicity of the elastic crystal and of the substrate. The ground-state phase diagram of the FK model has been extensively studied for nearest-neighbour interactions: When the corresponding ratio is an incommensurate number, at zero temperature the FK model reproduces the essential features of the stick-slip motion characteristic of static friction, with a continuous transition from sliding to pinning at finite lattice depths. As a function of the mismatch, the ground state is non-analytic and has a form of a devil's staircase, whose steps correspond to the regime of stability of a commensurate structure, i.e.\, a periodic structure pinned by the lattice \cite{Aubry:1983}. The transition to a sliding phase is characterized by the proliferation of kinks, namely, of local distributions of excess particles (or holes) in the substrate potential \cite{Braun_Kishvar}. Experiments with trapped ions observed several features of this dynamics: Stick-slip motion has been reported in chains of few ions \cite{gangloff_velocity_2015,Bylinskii:2015,Bylinskii:2016}, pinning by an external lattice has been observed \cite{Linnet:2012}, the onset of the Aubry transition has been measured in an implementation simulating a deformable substrate \cite{Kiethe:2017}, and the kinks density has been revealed in small chains as a function of the mismatch \cite{Gangloff:2022}.

These results show the versatility of trapped ion platforms as quantum simulators. Recent progress in cooling large ion chains \cite{Lechner:2016,Feng:2020} and loading ions in optical lattices \cite{Schmidt:2018,Hoenig:2023} pave the way towards studying kinks dynamics and their mutual interactions, thus shedding light into the interplay between geometric frustration and quantum fluctuations. In this regime, long-range forces, such as the Coulomb repulsion, qualitatively modify the kinks and the nature of their interactions \cite{Pokrovsky_1983}. A systematic study can be performed in the continuum limit, when the substrate potential is a small perturbation to the chain's interaction and the kinks are sine-Gordon solitons for nearest-neighbor interactions \cite{Merwe}. Then, the long-range interactions modify the sine-Gordon equation introducing an integral term \cite{Landa:2020} and the long-range sine-Gordon model can be mapped to an extended massive (1+1) Thirring Hamiltonian, where the solitons are charged positive-energy excitations over a Dirac sea \cite{Menu:2023}. This theory has a predictive power for ion chains provided the  {kink's size} is orders of magnitude larger than the  {interparticle distance}, allowing one to discard the discrete nature of the charge density distribution. The theory does not capture the opposite limit, where either the number of ions is limited to few dozens \cite{Bylinskii:2016,Kiethe:2018,Gangloff:2022} and/or the depth of the substrate potentials localizes the kinks in {chains composed by} few ions as in Refs.\ \cite{Landa:2013,Partner:2013}. In some treatments the discrete nature of the charge distribution 
can be theoretically described as 
corrections to the continuum limit \cite{ Willis:1986,Braun:1990,Chelpanova:2024}, leading to effective soliton-phonon collisions \cite{Willis:1986}. 

\begin{figure}
    \centering
    \includegraphics[width=0.9\columnwidth]{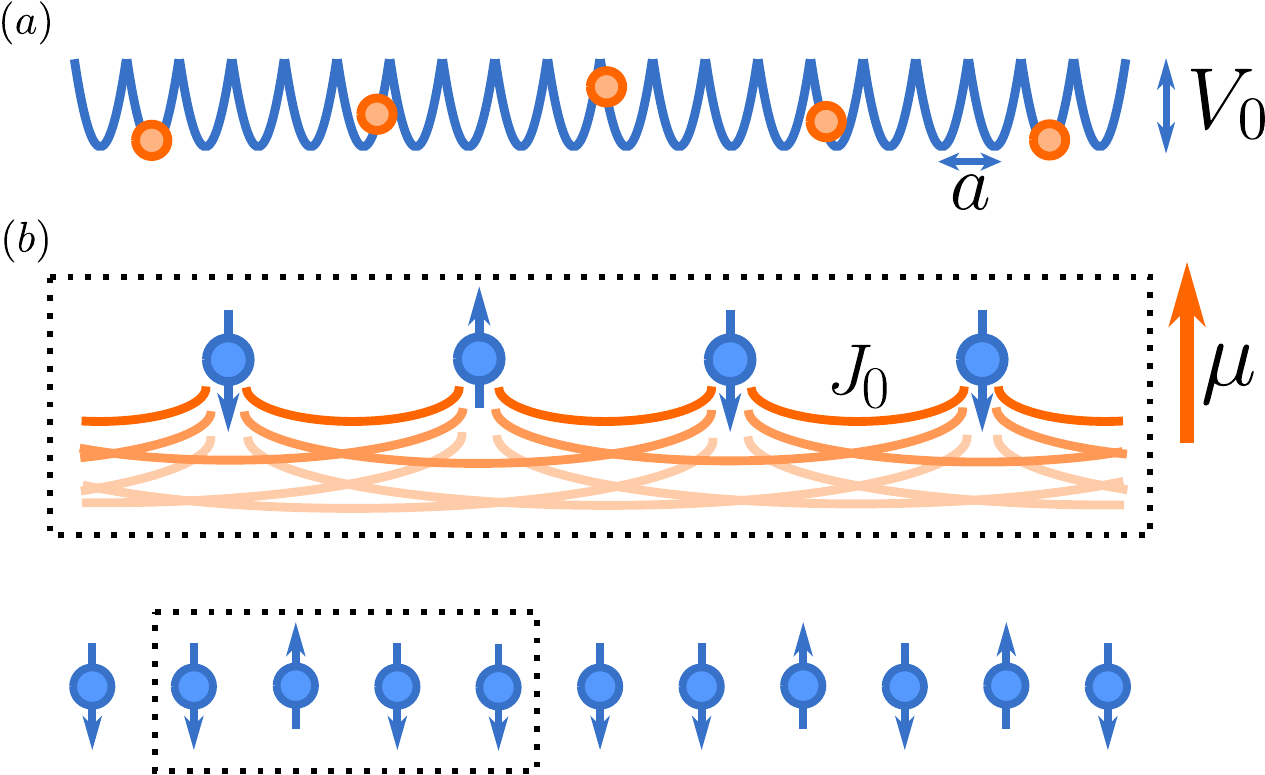}
    \caption{(a) Frenkel-Kontorova model where the periodic substrate potential is a piecewise parabolic function, Eq.\ \eqref{Eq:f} (in blue). The orange circles represent the interacting particles forming a chain. (b) The number of vacant sites $h_j$ between neighbouring particles is mapped onto a spin whose orientation depends on the value of $h_j$, as shown in Sec.\ \ref{Sec:3}. In the effective model, the depth of the substrate potential $V_0$ determines the short-range behavior of the antiferromagnetic spin-spin coupling $J_0$ while the mismatch is a magnetic field $\mu$. The central panel is a magnified fragment of the spin larger chain indicated by the dotted box. The power-law decay of the interactions is indicated by the gradual attenuation of the connecting edges. }
    \label{fig:2}
\end{figure}

In the present work, we choose a different approach and start from a discrete distribution of interacting particles. Due to the long-range interactions, the ground state emerging from the competition of interactions and substrate potential cannot be found by means of the ingenious dynamical map of Ref. \cite{Aubry:1983}. We instead implement the method of Hubbard \cite{Hubbard}, and map the ground state configuration of the long-range, Coulomb Frenkel-Kontorova model to the one of an antiferromagnetic spin chain in the presence of a magnetic field. The mapping is exact for a periodic substrate composed of piecewise harmonic oscillators \cite{Pietronero}, illustrated in the upper panel of Fig.\ \ref{fig:2}, and is amenable to analytical solutions. Despite the theoretical abstraction, we show that this mapping sheds light on the properties of realistic substrate potentials, such as an optical lattice.

The presentation of our study is organised as follows. In Sec. \ref{Sec:1} it is shown that the ground state and low-energy excitations of a Wigner crystal of ions in a linear Paul trap are described by a  Frenkel-Kontorova (FK) model where the oscillators of the elastic crystal interact via the long-range Coulomb interactions. This Section reviews the arguments presented in Refs. \cite{Garcia-Mata,Pruttivarasin:2011,Cormick:2013} and sets the stage for our analysis. The ground state is determined in Sec. \ref{Sec:2} within a mean-field approach, which discards the kinetic energy. Here, we assume a specific function of the periodic substrate and map the continuous-variable problem onto an  {long-range} Ising model and in the presence of a magnetic field, as illustrated in the lower panel of Fig.\ \ref{fig:2}. Our mapping extends the study of Ref. \cite{Pietronero} to a Wigner crystal and allows us to show that the ground state is a devil's staircase as a function of the mismatch between the lattice periodicity and the characteristic interparticle distance. It allows us, moreover, to determine the interval of stability of the individual commensurate structures as a function of the temperature. In Sec. \ref{Sec:3} we determine the low-energy excitations of an ion chain in a  {sinusoidal} potential across the Aubry transition and identify its experimental signatures. We then discuss the order of magnitude of quantum fluctuations by means of a semiclassical ansatz. The conclusions are drawn in Sec. \ref{Sec:5}, where we provide an outlook of the directions of studies that our work opens towards the systematic characterization of the interplay between long-range interactions and geometric frustration with cold atoms platforms.

\section{Chain of interacting particles in a periodic potential}
\label{Sec:1}

This section reviews the basic assumptions and the steps that connect the Hamiltonian of a one-dimensional Wigner crystal of ions in a periodic potential with a Frenkel-Kontorova model of oscillators interacting via power-law decaying forces.  We then generally discuss the geometric properties of the ground state using the characterization of Hubbard \cite{Hubbard} and introduce the quantities that will be important for performing the mapping in Sec. \ref{Sec:2}. We refer the interested reader to Refs. \cite{Garcia-Mata,Pruttivarasin:2011}, where it was proposed to study the sliding-to-pinning transition using Wigner crystals of trapped ions in optical lattices. 

\subsection{Extended Frenkel-Kontorova model}

We consider $N$ particles of mass $M$ in one dimension, ordered along the $x$--axis. Let $x_j$ be the particles' positions, with $j=1,\ldots, N$, such that $x_j<x_i$ for $j<i$. We denote by $L$ the chain's length and assume periodic boundary conditions. The particles interact via the repulsive two-body potential $W(x)$, that decays algebraically with the distance $x$ as $$W(x)=W_0/x^\alpha\,$$ with $W_0>0$. In this section we keep the power law exponent $\alpha$ generic, restricting to values $\alpha\ge 1$, hence including also the Coulomb interaction. 

The overall potential energy includes a periodic substrate potential $V_s(x)$ and takes the form
\begin{equation}
\label{Eq:V:0}
V = \dfrac{W_0}{2}\sum_{i,j}\dfrac{1}{\vert x_i -x_j\vert^\alpha} + \sum_j V_s(x_j),
\end{equation}
where we assume periodic boundary conditions and that $V_s(x)$ has periodicity $a$, $V_s(x+a)=V_s(x)$. For later convenience, we write the substrate potential as $$V_s(x)=V_0f(x),$$ where $V_0\in\mathbb{R}^+$ determines the depth of the potential and $f(x)$ is a dimensionless periodic function with unit amplitude. 

In order to link the model of Eq. \eqref{Eq:V:0} with the paradigmatic Frenkel-Kontorova model, we assume that the particles are localized about the equilibrium positions of the potential $W(x)$. We perform a Taylor expansion of the interaction $W(x)$ about the classical equilibrium positions $x_j^{(0)}$ assuming that the average interparticle distance $d_0=L/N$ is much larger than the lattice periodicity $a$, thus $x_j^{(0)}=jd_0$. We denote by $u_j$ the local displacement of the particle $j$ from the equilibrium position $x_j^{(0)}$, such that $x_j = x_j^{(0)} + u_j$. In second order in the expansion in the small parameter $u_j$ ($u_j\ll d_0$) the potential reads:
\begin{equation}
\label{Eq:V:Taylor}
    V \simeq W^{(0)} + V_0\sum_j{f(x_j)} + \dfrac{1}{2}\sum_j\sum_{n>0}{\dfrac{K}{n^{\alpha + 2}}(u_{j+n} - u_j)^2}\,,
\end{equation}
where $W^{(0)}$ is the interaction potential at the equilibrium positions, $$W^{(0)}=\frac{W_0}{2}\sum_{i,j}\dfrac{1}{\vert x_i^{(0)} -x_j^{(0)}\vert^\alpha}\,,$$ and $K$ is the spring stiffness, 
\begin{eqnarray*}
K = \frac{\alpha(\alpha + 1)W_0}{d_0^{\alpha+2}}\,.
\end{eqnarray*}

\noindent Equation \eqref{Eq:V:Taylor} corresponds to the potential of the Frenkel-Kontorova model with power-law elastic interactions.

Some words of caution about this treatment shall be spent. In fact, the validity of the Taylor expansion requires that the classical ground state is stable against fluctuations. In one dimension this is not verified for interactions with exponent $\alpha>1$: In that case the treatment here presented is valid only for sufficiently small chains, while for long chains the ground state is captured by a Luttinger model, see Ref.\ \cite{Dalmonte:2010}. The Coulomb chain, $\alpha=1$, is a special case due to the non-additivity of the energy, that leads to the slow decay of two-point density correlations with distance \cite{Schulz:1993,Morigi:2004}. As a consequence, for any finite size the Coulomb chain exhibits long-range order even at zero temperatures. 

\subsection{Potential of the vacant sites}

Hereafter, we will assume that at most one particle is assigned to each lattice site. In order to distinguish classical configurations, we will introduce the notation of Ref.\ \cite{Pietronero}: Let $h_n$ be the number of vacant sites between two subsequent particles of the chain. The sequence $\lbrace h_1, \dots, h_N \rbrace$  fully characterizes a classical equilibrium configuration. The potential energy of Eq. \eqref{Eq:V:Taylor} can be expressed in terms of the sequence of vacant sites, $\lbrace h_1, \dots, h_N \rbrace$, via the equivalent reformulation of the position variables
\begin{equation}
\label{Eq:h:0}
    x_j = a\sum_{i=1}^{j-1}{(h_i + 1)} + \delta x_j,
\end{equation}
where now $\delta x_j$ is the displacement of the particle with respect to the closest substrate-potential well. Using Eq.\ \eqref{Eq:h:0} and that $f(x_j) = f(\delta x_j)$, we cast the potential, Eq. \eqref{Eq:V:Taylor}, in the form: 
\begin{widetext}
\begin{align}
\label{Eq:V:h}
    V &\simeq V^{(0)} + V_0\sum_{j}{f(\delta x_j)}+\dfrac{K}{2}\sum_{j}\sum_{n>0}{\dfrac{1}{n^{\alpha + 2}}(\delta x_{j+n} - \delta x_j)^2}
    + aK\sum_{j}\sum_{n>0}{\dfrac{1}{n^{\alpha + 2}}(\delta x_{j+n} - \delta x_j)(h^{(n)}_j - n\langle h \rangle)}\notag\\
    &+a^2\dfrac{K}{2}\sum_{j}\sum_{n>0}{\dfrac{1}{n^{\alpha + 2}}(h^{(n)}_j - n\langle h \rangle)^2},
\end{align}
\end{widetext}
where $V^{(0)}$ is the potential in zeroth order in the expansion in $\delta x_j$, and we have introduced the notation 
\begin{equation}
h^{(n)}_j = \sum_{i=0}^{n-1}{h_{j+i}}\,.\label{Eq:h:n}     
\end{equation}
Equation \eqref{Eq:V:h} is the starting point for performing a mapping to a potential of interacting spins. A crucial part of this mapping consists in eliminating the displacement variables $\delta x_j$ and rewriting the potential energy only in terms of the vacant-site variables $h_n$, which in turn will be mapped onto spins.

\subsection{Equilibrium configurations}

The ground states configurations of potential \eqref{Eq:V:h} are determined by the competition of the power-law interaction and the periodic substrate potential. Moreover, they shall satisfy the additional constraint of periodic boundary conditions. We first note that the length $L$ of the chain shall be an integer multiple $N_s$ of the substrate periodicity $a$: $L=N_sa$. This establishes a relation between the average interparticle distance, $d_0=L/N$, and the lattice periodicity $a$, given by $N d_0 = N_s a$. From these quantities we find the mean number of particles per lattice site, which we denote by $\rho$:
\begin{equation}
    \rho = \dfrac{N}{N_s} = \dfrac{a}{d_0}\,.
\end{equation}
We can further link the density with the the average number of empty sites, $\langle h \rangle = \sum_j h_j/N$, by observing that the sum of vacant sites shall fulfil the relation
\begin{equation}
    \sum_j h_j = N_s - N\,.\label{Eq:h}
\end{equation}
Dividing both sides by $N$, we link the average number of empty sites with the density of charges:
\begin{equation}
    \langle h \rangle = \dfrac{1}{\rho} - 1\,.\label{h:mean}
\end{equation}

Due to the periodic boundary conditions, the structures emerging from the competition between the substrate potential and the two-body interactions are necessarily periodic. True incommensurate structures will then exist in the strict thermodynamic limit. For finite-size chains we will denote a structure as {\it incommensurate} when the following condition occurs. Let $P$ be the period characterizing the structure: $P=L/M$ with $M$, a natural number such that $M\ge 1$. A structure will be commensurate when $M>1$. On the contrary, incommensurate configurations are characterized by $P =L$, namely, the period is the full length of the chain. See also Ref.\ \cite{Roux:2008} for a related discussion.

In what follows we will consider the case $\rho<1$. The Taylor expansion of Eq.\ \eqref{Eq:V:Taylor}, in particular, requires that the particles displacement is of the order of the lattice periodicity and thus is valid for densities $\rho\ll 1$.

\section{Ground state of the piece-wise parabolic potential}
\label{Sec:2}

\begin{figure*}
    \centering
    \includegraphics[width=0.9\textwidth]{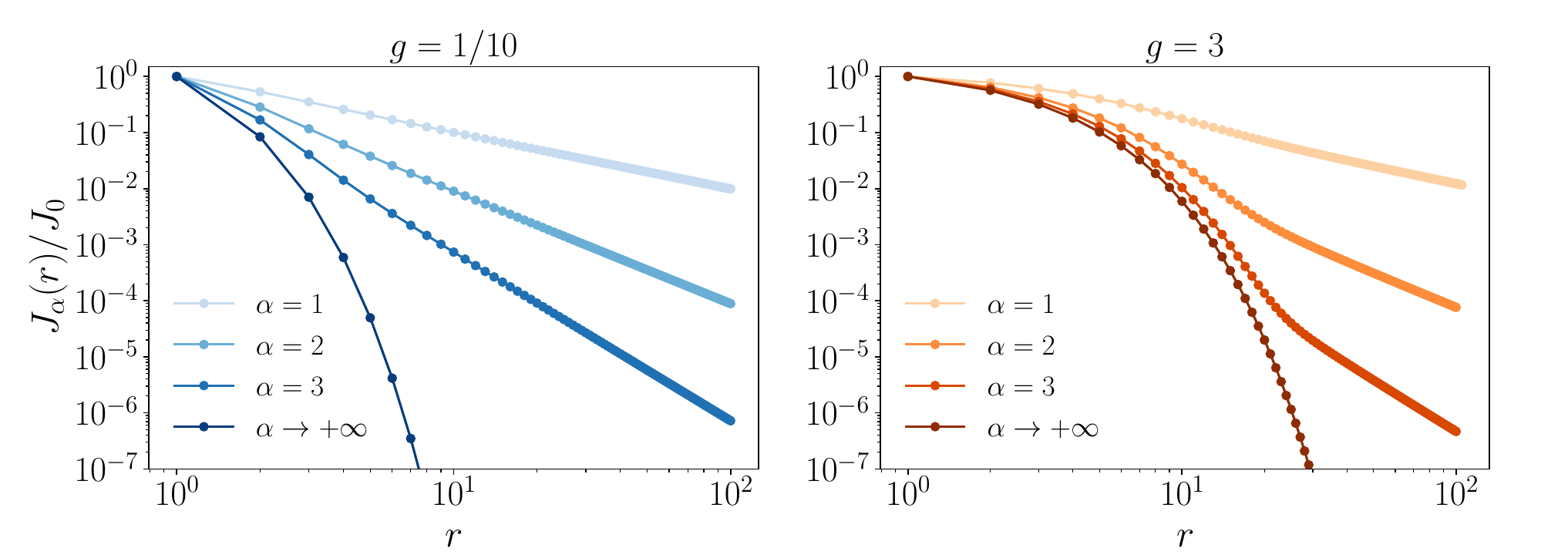}
        \caption{{Scaling of the} coefficients of the interacting potential, Eq.\ \eqref{Eq:MF}, {with the distance $r$}. The coefficient $J_\alpha(r)$, Eq.\ \eqref{eq:J:0} are displayed as a function of $r$ (in units of $a$) for a deep ($g=1/10$) and a shallow ($g=3$) substrate lattice. The colors refer to different values of the power-law exponent $\alpha$, see legenda. The logarithmic scale emphasizes the two-fold behavior of the coefficients, which at short distances decay exponentially while at long distances decay algebraically as $1/r^\alpha$ for finite exponents $\alpha$. The nearest-neighbor case ($\alpha\to\infty$) is solely characterized by the exponential decay.}
    \label{fig:3}
\end{figure*}

We now show that the model of Eq.\ \eqref{Eq:V:h} can be mapped onto a long-range  Ising model in the presence of a magnetic field, as illustrated in Fig.\ \ref{fig:2}(b). The classical ground state of this model is a devil's staircase as a function of the density $\rho$ \cite{Bak-PRL:1982}. The mapping we perform is exact  when the periodic substrate potential is a sequence of piecewise-truncated parabolas of the form 
\begin{equation}
\label{Eq:f}
f(\delta x) = \dfrac{4}{a^2}\delta x^2,
\end{equation}
for $\vert \delta x \vert \leq a/2$, see Fig.~\ref{fig:2} (a). This functional behavior 
has been used in several analyses (see, e.g., \cite{Aubry:1983a,Pietronero,Mueser:2005}).

\subsection{Mean-field configuration in Fourier representation}

The mapping is performed by first eliminating the displacement $\delta x_j$ from the potential of Eq. \eqref{Eq:V:h} and expressing the potential itself as a function of the segments of vacancies, $h_j$. For this purpose, we introduce the Fourier components for the variables of interest
\begin{subequations}
\begin{align}
    Q_q &= \dfrac{1}{\sqrt{N}}\sum_j{e^{-iqja}\delta x_j},\\
    \zeta_q &= \dfrac{1}{\sqrt{N}}\sum_j{e^{-iqja}(h_j - \langle h \rangle)}\,,
\end{align}
\end{subequations}
where $q$ is the wave number in the Brillouin zone of the lattice. For convenience, we also introduce the Fourier components of the sequences of vacancies, $h^{(n)}_j$, namely
\begin{align}
    \zeta^{(n)}_q &= \dfrac{1}{\sqrt{N}}\sum_j{e^{-iqja}(h^{(n)}_j - n\langle h \rangle)}\,.\notag
\end{align}
Using Eq. \eqref{Eq:h:n}, this expression takes the compact form:
\begin{align}
\zeta^{(n)}_q&= \zeta_q \sum_{j=0}^{n-1}{e^{iq j a}}
=\dfrac{1 - e^{iqna}}{1-e^{iqa}}\zeta_q\,.
\end{align}
On the basis of these definitions, the potential energy can be rewritten in terms of the Fourier components.
\begin{align}
    V &= V^{(0)} + \dfrac{8V_0}{a^2}\sum_{q>0}{Q_q Q_{-q} (1 + g\phi_\alpha(q))}\notag \\
    & + a\dfrac{8V_0}{a^2}\sum_{q>0}\left({Q_q \zeta_{-q} \dfrac{g \phi_\alpha(q)}{e^{-iqa} - 1} + \mathrm{c.c.}}\right)\notag \\
    & + a^2\dfrac{8V_0}{a^2}\sum_{q>0}{\zeta_q \zeta_{-q} \dfrac{g \phi_\alpha(q)}{\vert e^{iqa} - 1 \vert^2}},
\end{align}
where the long-range nature of the interactions is now encoded in the function $\phi(q)$, defined such that
\begin{equation}
\label{Eq:Phi}
    \phi_\alpha(q) = \sum_{n>0}{\dfrac{\vert 1 - e^{iqna} \vert^2}{n^{\alpha + 2}}}\,.
\end{equation}
Now, the dimensionless coefficient 
\begin{equation}
\label{Eq:g}
 g = \frac{K a^2}{8 V_0}   
\end{equation}
quantifies the competition between the elastic properties of the chain and the interaction with the substrate.

For nearest-neighbor interactions ($\alpha\to\infty$) the parameter $g$ controls the transition from sliding, where {kinks} proliferate ($g\gg 1$), to pinning ($g\ll 1$), where the formation of solitons is energetically costly. 

Our interest lies in determining how the periodic potential stabilizes a new ordering of the chain of particles. Equilibrium requires the condition $\frac{\partial V}{\partial Q_q} = 0$  for all values of the wave number $q$. This condition leads to a linear relation between the displacements and the sequences of vacant sites \footnote{We note that, in the nearest-neighbour limit $\alpha\to\infty$, this expression reduces to Eq.\ (3.17) of Ref. \cite{Pietronero}. In order to perform a systematic comparison, we note that the coefficient $A$ of Ref. \cite{Pietronero} corresponds to our coefficient $\mathcal K$ and their coefficient $g$ is twice the coefficient $g$ of Eq. \eqref{Eq:g}. With these substitutions Eq. \eqref{Eq:Qq:1} coincides with Eq. (3.17) of Ref. \cite{Pietronero}.}
\begin{equation}
    Q_q = -a\,g \dfrac{\phi_\alpha(q)}{1 + g \phi_\alpha(q)}\left(\dfrac{1}{e^{iqa} - 1} \right)\zeta_q \,\label{Eq:Qq:1}.
\end{equation}
We note that the coefficient $g$ is inversely proportional to the square of the mass of the sine-Gordon kink $M_{\rm SG}$ \cite{Braun_Kishvar,Landa:2020}: 
{\begin{equation}
    M^2_{\rm SG}=\dfrac{\pi^2}{3g}\,.
\end{equation}}

\subsection{The potential for the vacant sites}

By means of Eq.~\eqref{Eq:Qq:1}, we can recast the expression of the potential energy in terms of the Fourier components of the vacant sites only. In real space, the potential for the vacant sites takes the form:
\begin{eqnarray}
\label{Eq:MF}
V=V^0+\sum_{\ell}\sum_rJ_\alpha(r)( h_{\ell}-\langle h\rangle)(h_{\ell+r}-\langle h\rangle)\,,
\label{eq:Vint}
\end{eqnarray}
with the interaction coefficient:
\begin{eqnarray}
J_\alpha(r)&=& \frac{4V_0}{N}\sum_{q}\cos(qra)\frac{g}{1+g\phi_\alpha(q)}\,\frac{\phi_\alpha(q)}{\phi_\infty(q)}\,,\nonumber\\
\label{eq:J:0}
\end{eqnarray}
where $\phi_\infty(q)=\lim_{\alpha\to\infty} \phi_{\alpha}(q)=\vert e^{iqa} - 1\vert ^2$.

It is instructive to analyse the generic behavior of the coefficients $J_\alpha(r)$ as a function of $r$ for finite power-law exponents $\alpha$, see Fig.~\ref{fig:3}. We first note that $|\phi_\alpha(q)|\le 2\zeta(\alpha+2)$, with $\zeta(\alpha)$ the Riemann zeta function. By means of an analytic continuation, it becomes visible that the pole of the function $\mathcal F_\alpha(q)=1+g\phi_\alpha(q)$ determines an exponentially decaying behavior with a characteristic length that is monotonically increasing with $g$, see Appendix~\ref{App:A}. For nearest-neighbor interactions ($\alpha \to +\infty$), the coefficient decays exponentially with a damping length monotonically increasing with $g$. Instead, for finite values of $\alpha$, we observe a two-fold behaviour of $J_\alpha(r)$: at short distances the coefficient decays exponentially, {with a characteristic length depending on $g$,} whereas at long distances $|r|\gg a$ the coefficient exhibits a power-law tail solely determined by the long-range interactions. In Appendix~\ref{App:A} we show that the power-law tail takes the form
\begin{equation}
    J_\alpha(r) \simeq \dfrac{K a^2}{2}\dfrac{\Gamma(\alpha)}{\Gamma(\alpha + 2)}\dfrac{1}{\vert r \vert^\alpha}\,\label{Eq:J},
\end{equation}
which is independent of $g$. At large distances, thus, the coefficient describes a power-law repulsion at the same exponent $\alpha$ of the interaction. This is in agreement with the general considerations of Refs.\ \cite{Pokrovsky_1983,Braun:1990,Landa:2020}. The short-distance and large-distance behavior of the coefficient $J_\alpha(r)$ is visible in Fig.\ \ref{fig:3} for deep ($g=1/10$) and shallow lattices ($g=3$) for representative values of the exponent $\alpha$.%

\subsection{The dislocation}

\begin{figure}
    \centering
    \includegraphics[width=0.9\columnwidth]{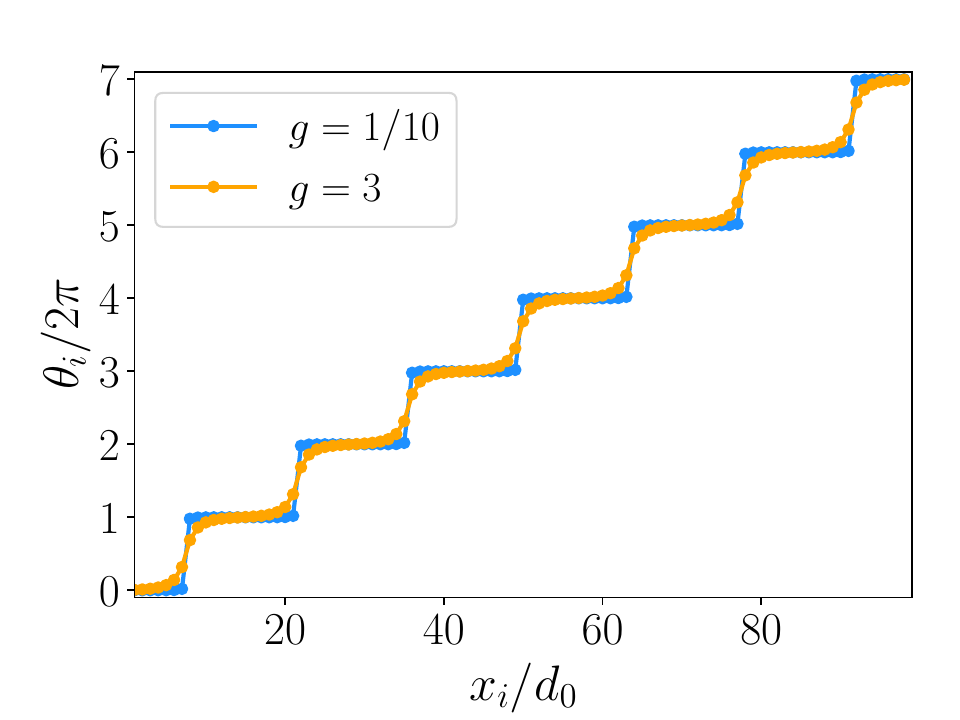}
    \caption{Dislocations for different lattice depths. A phason $\theta_i/2\pi$, corresponding to the in-well displacement $\delta x_j/a$, is displayed as a function of the ion position $x_i$. Calculations are performed for $N=99$ ions and density $\rho = 99/700$ in the case of a deep (blue) and shallow (orange) substrate potential.} 
    \label{fig:4}
\end{figure}

By transforming back into the space variables, we obtain the equilibrium positions of the ions as a function of the empty sequences. For $g\gg 1$ the displacements take the form (see Appendix \ref{App:B}:)
\begin{eqnarray}
\delta x_j&\approx& a\frac{g}{\alpha+1}\sum_{r>0}\frac{1}{r^{\alpha+1}}(h_{j+r}-h_{j-r})\,.\label{eq:delta:x:0}
\end{eqnarray}
This expression provides the shape of the dislocation. We introduce the phase field $\theta_j/2\pi$:
\begin{equation}
    \theta_j=\frac{2\pi}{a}\left(x_j - ja\left(\dfrac{d_0}{a}-\delta\right)\right),
\end{equation}
The phase field is displayed on Fig.~\ref{fig:4} for two values of the coefficient $g$. Each step is a dislocation inside the ion chain. Decreasing the value of $g$, thus increasing the amplitude of the substrate potential, leads to increasingly sharper jumps in the shape of the phason, as the ions become pinned to the local minima of the substrate potential.

\subsection{A long-range {Ising} model}
\label{Sec:3D}

The segments $h_j$ in Eq.\ \eqref{Eq:MF} can be interpreted as interacting spins \cite{Pietronero}. For this purpose, it is now useful to recall that the segments of vacant sites $h_j$ can only be integer numbers. In a commensurate structure where the equilibrium interparticle distance is $\bar d_0 = (n_0 + 1)a$, the number of vacancies is uniform and equal to $h_j = n_0$. 
A discommensurate structure, instead, is characterized by an average interparticle distance 
\begin{equation}
d_0 = \bar d_0 + \delta a\,,
\end{equation} 
where the  parameter $\delta$ determines the discommensuration (or mismatch), $0<\delta <1$. In the ground state the segments rearrange such that $h_j$ can either be $n_0$ or $n_0 + 1$, satisfying the constraint imposed by Eq.~\eqref{Eq:h}, see Refs.\ \cite{Hubbard,Bak-PRL:1982}. Since the number of vacant sites can only take two values, we will treat them as classical spins $S_j=\pm 1$ where $$S_j=2(h_j-n_0)-1\,.$$ Thus, $S_j=1$ when $h_j= n_0 +1$ and $S_j=-1$ when $h_j= n_0$. We use that $\langle h \rangle = n_0 + \delta$ and rewrite the potential energy as a spin Hamiltonian of the form
\begin{equation}
    {\mathcal H} = \dfrac{1}{4}\sum_{i\neq j}{J(i-j)(S_i + 1)(S_j + 1)} - \mu\sum_i{S_i}\,,
    \label{H:BB}
\end{equation}
where $J(r)=J_\alpha(r)$. { The first term of this Hamiltonian describes an antiferromagnetic Ising chain in a polarizing magnetic field of strength $\sum_{r>0}J(r)$. This polarizing field is shifted by an offset}
\begin{equation}
\label{Eq:mu}
\mu =2\delta\sum_{r>0}J(r)\,.
\end{equation}
{ This additional} magnetic field, in turn, is proportional to the discommensuration and tends to align the spin, competing with the antiferromagnetic order imposed by the interactions.

The Hamiltonian in Eq.~\eqref{H:BB} is given apart for a constant energy offset $E_0$, which is positive and depends on the discommensuration: $E_0=2N\delta(\delta-1/2)\sum_{r>0}J(r)$.
\subsection{Devil's staircase}

\begin{figure}
    \centering
    \includegraphics[width=0.9\columnwidth]{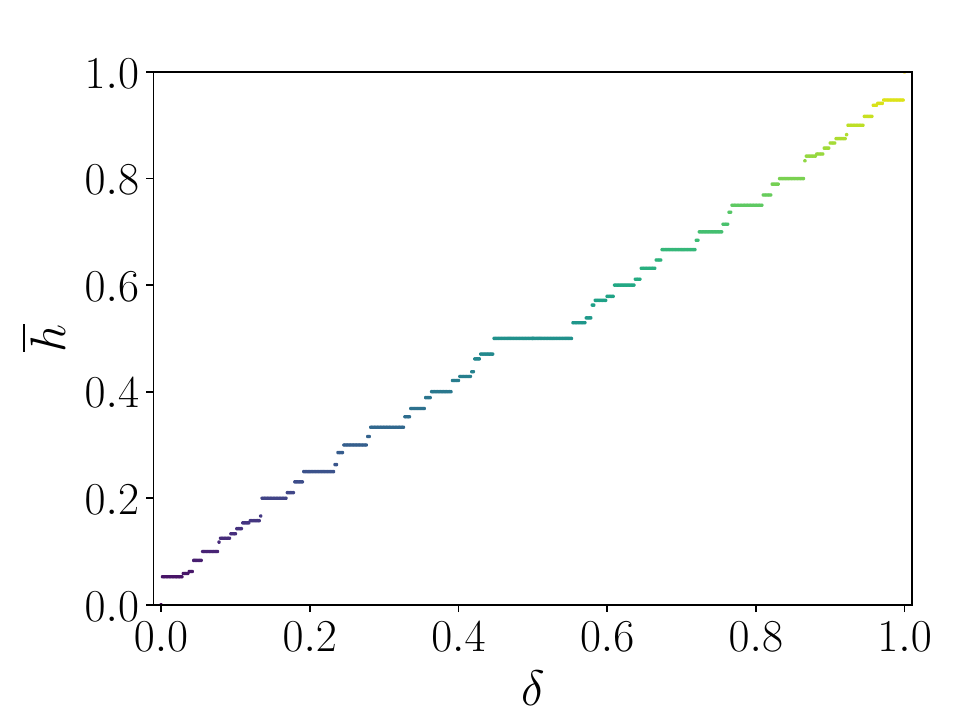}
    \caption{Devil's staircase of commensurate structures as a function of the mismatch. The plot shows the magnetization, here represented in terms of the ratio $\overline{h}=m/n$ of up-oriented spins, as a function of the magnetic field, here given by the discommensuration $\delta$. The staircase has been numerically determined using the method of Ref.\ \cite{Koziol:2023} for a chain of $N=200$ ions ($\alpha = 1$) and {$g = 1/10$.} The displayed plateaus correspond here to the ratios $m/n$ with $n\leq 20$. } 
    \label{fig:5}
\end{figure}

The parameters $\delta$ and $g$ fully characterize the classical ground state. For nearest-neighbour interactions, the phase diagram of the spin model of Eq. \eqref{H:BB} entails the so-called Aubry transition, where, at fixed mismatch $\delta$, a critical depth of the potential $V_0$ separates the sliding phase from the pinned phase. Here, the spectrum is gapped and the phase is dynamically characterized by stick-slip events. At constant $g$, the phase diagram also entails the so-called commensurate-incommensurate transition: here a critical value of the mismatch separates an ordered (commensurate) phase from the sliding phase where kinks proliferate \cite{Bak_1982}. At both transitions the ground state becomes non-analytic. The fractal nature of the ground state becomes visible when considering the so-called magnetization $\overline{m}$ as a function of the magnetic field (and thus of the mismatch). The magnetization $\overline{m}$ is defined as
\begin{equation}
\overline{m}=\sum S_j/N=2\bar h-1\,,
\end{equation} 
with $\bar h=\langle h\rangle -n_0$ the effective discommensuration. In the absence of the substrate potential, $\bar h=\delta$, and the magnetization is proportional to the magnetic field. At finite substrate depths instead, $\bar h$ exhibits a devil staircase as a function of $\delta$. {The staircase exists for all values of $\alpha>1$ \cite{Bak-PRL:1982}. The Coulomb case $\alpha=1$ {is discussed in the next section.} 
\begin{figure*}
    \centering
    \includegraphics[width=0.8\textwidth]{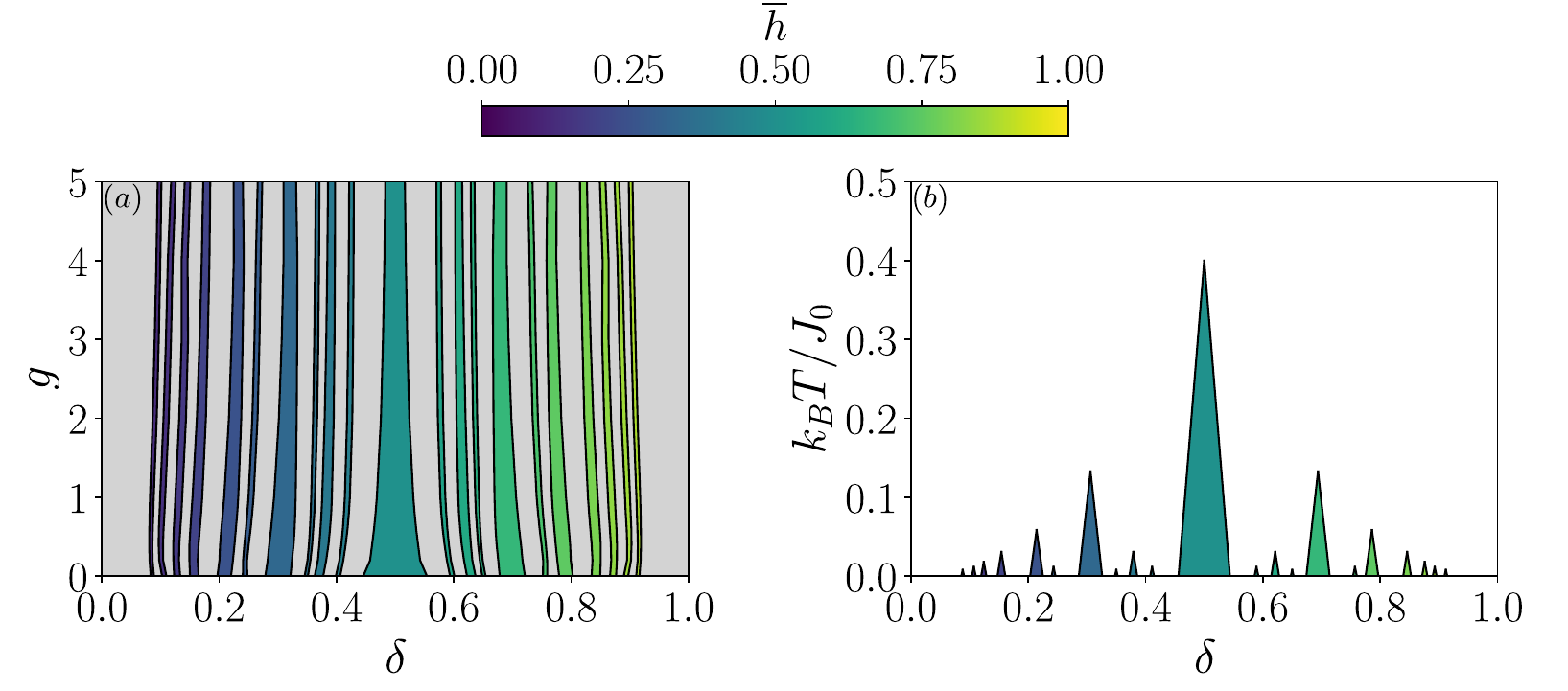}
    \caption{(a) Magnetization, $\overline{h}=m/n$, as a function of the discommensuration $\delta$ and of the ratio $g$ between the stiffness and the potential depth. Colored areas represent plateaus corresponding to ratios $\overline{h}$ with $n\leq 8$. In the {grey shaded} areas the plateaus of the staircase are below the chosen resolution. (b) The magnetization $\overline{h}=m/n$ as a function of the discommensuration $\delta$ and of the temperature $T$ for {$g=1/10$}. Colored areas represent plateaus corresponding to ratios $\overline{h}$ with $n\leq 8$. White areas correspond to a paramagnetic, disordered phase. All calculations are performed for a chain of $N=200$ ions interacting via Coulomb repulsion ($\alpha = 1$).}
    \label{fig:6}
\end{figure*}

\subsection{Thermodynamic limit and phase transitions}

 {Figure\ \ref{fig:5} displays a staircase for a finite chain of ions at a fixed, small value of $g $. The 
{stability regions in the $g-\delta$ plane of some commensurate phases are shown} in Fig.~\ref{fig:6}(a). One observes that the size decreases as the lattice depth decreases (corresponding to increasing $g$).}
 {The size of the steps of the devil's staircase as a function of $\delta$} agrees with an analytical expression obtained by means of sum rules for generic, convex interactions \cite{Hubbard,Bak-PRL:1982}. For a magnetization with $\bar h=m/n$ ( $m$ and $n$ natural numbers and prime to each other) the interval of stability is given by \cite{Bak-PRL:1982} 
\begin{eqnarray}
\Delta \mu_\alpha\left[\dfrac{m}{n}\right] = \sum_{j=1}^{N}{n j \left[J_\alpha(j n -1) + J_\alpha( j n + 1) - 2J_\alpha(j n) \right]}\,,
\label{Eq:Bak}
\end{eqnarray}
 {and does not depend on $m$.}
This analytical expression provides the boundaries of the stability of classical commensurate structures in the $g-\delta$ plane.  {It is the energy gap for flipping a spin in the commensurate phase and thus the energy for generating a kink.} When the gap vanishes, flipping a spin (generating a dislocation) becomes energetically favorable and kinks proliferate.  {At fixed mismatch $\delta$, the gap vanishes at the critical value $g_c(\delta)$, determining the Aubry transition separating a sliding (gapless) from a pinned (gapped) phase.
At fixed $g$, the mismatch $\delta_c(g)$ is the critical value at which the commensurate-incommensurate transition occurs. }

 {For $\alpha>1$, by means} of a proper rescaling (Kac's rescaling) \cite{CAMPA200957,Defenu:2019}, the critical values $g_c$ and $\delta_c$ tend to a finite value in the thermodynamic limit $N\to\infty$ . For $\alpha=1$, instead, the {steps of the staircase} vanish.
This is a consequence of the non-additive nature of the energy for Coulomb interactions in one dimension, {which scales as $N\ln N$ and tends thus to dominate over the substrate potential as $N\to\infty$. The vanishing of $\Delta\mu$ as $N\to \infty$ can be illustrated by renormalizing the charge as $Q^2\to Q^2/\ln N$ \cite{Morigi:2004,Landa:2020}. With this rescaling, $g\propto 1/\ln N$, the interaction coefficients scale as $1/\ln N$ and correspondingly the size of the plateaus $\Delta\mu\sim 1/\ln N$. Therefore, $\Delta\mu\to 0$ for $N\to\infty$ and the staircase disappears in the thermodynamic limit.} 

This behavior is in agreement with the prediction that for Coulomb interactions the fractal dimension is unity \cite{Bruinsma:1983}. It is a manifestation of the long-range nature of the Coulomb interaction which tends to prevail over the order imposed by the external lattice. As a consequence, the Aubry and commensurate-incommensurate transitions are {crossovers for all $N$}. Nevertheless, given the extremely slow growth of $\ln N$ with $N$, a {pinned or a commensurate phase} can be experimentally measured for any chain size, provided that the temperature of the chain is sufficiently low, as we quantify in the next section.

\subsection{Thermal effects}

With an argument based on the scaling of entropy in the free energy, Peierls showed that in one dimension thermal fluctuations prevent the emergence of magnetic order \cite{Peierls}. This observation holds in the thermodynamic limit and for systems with additive energy. For finite systems, there is a temperature $T(N)$ above which the commensurate structure becomes unstable. The temperature $T(N)$ decreases with $N$, and vanishes in the thermodynamic limit. 

We estimate $T(N)$ using a semiclassical model, where we calculate the change of free energy by creating a defect in the commensurate structure as $\Delta F=\Delta E-T\Delta S$, where $\Delta E$ is the change in energy and $\Delta S$ the one in entropy. By means of the mapping to the antiferromagnetic spin model, then $\Delta E=\Delta\mu$ of Eq. \eqref{Eq:Bak}. The change in entropy can be determined within the spin model. For a $n$-partite ordered magnetic phase, the total entropy takes the form \cite{PhysRevLett.126.183601}
\begin{equation}
    S = \dfrac{N k_B}{n}\left[ n\ln 2 - \sum_{\sigma = \pm}\sum_{r=1}^{n}{\dfrac{1 + \sigma m_r}{2}\ln(1 + \sigma m_r) } \right],
\end{equation}
where the set $\lbrace m_r\rbrace$ corresponds to the magnetization of each of the $n$ sublattices. For a perfectly ordered phase ($m_r = \pm 1$), the entropy cost of flipping a single spin (so for a variation of magnetization $d m_r = n/N$) scales like in the thermodynamic limit as $S \sim k_B\ln (N/n)/2$. Therefore, the free energy cost of flipping a spin starting from the magnetically ordered (commensurate) phase at a given value of $\overline h$ is given by
\begin{equation}
    \Delta F(\overline{h})=\Delta\mu[\overline{h}]- k_B T \dfrac{n}{4}\ln \left( \dfrac{N}{n}\right)\label{Eq:T},
\end{equation}
and it is stable for $\Delta F>0$. The quantity $\Delta F(\overline{h})$ provides the size of the plateaus at finite temperatures. Interestingly, also the entropy change depends on $n$ and increases with $n$. This expression also shows that the temperature below which the commensurate structure $\overline h$ is stable, scales as 
$$k_BT(N)\sim\frac{4\Delta_\mu(\overline{h})}{n}\frac{1}{\ln (N/n)}\,. $$

On the basis of this expression, {we determine the stability of the commensurate phase
with regard to thermal fluctuations, which we plot in Fig.~\ref{fig:6}(b) in the $\delta-T$ plane.}
We observe the progressive shrinking of the plateaus of the staircase as the thermal fluctuations become increasingly prominent. These results also allow to estimate the temperature below which one can expect to observe an incomplete devil's staircase in a realistic trapped-ion experiments. For an experimental set-up similar to the one realized in \cite{Bylinskii:2016}, one can expect to observe plateaus for temperatures $T$ below $T \lesssim 1 \, \mathrm{mK}$.

\subsection{Discussion}
{The phenomena featured in this section have been derived under the assumption that the substrate potential takes the form of a piece-wise set of parabola, which enables the exact mapping to a Ising model. While this might seem to be strictly valid when the substrate potential has the specific, discontinuous form, yet an analysis performed  with a substrate potential of sinusoidal form leads, when the kink size extends to several lattice sites, to the mapping to a XXZ Ising model \cite{Menu:2023}. In the classical limit, this model coincides with the Ising model of Eq.\ \eqref{Eq:Bak}, except in the specific form of the coefficients \cite{Menu:2023}. This suggests that the prediction of the piece-wise set of parabola can extend well beyond this specific substrate potential. One point of concern when assuming this potential are the cusps separating the wells. Nevertheless, we expect that the Ising model of Eq.\ \eqref{Eq:Bak} permits to estimate the stability region of commensurate phases in other continuous substrate potentials that are quadratic at the minima of the wells. In the next section we discuss an experimentally relevant substrate potential, the sinusoidal potential, and analyse the impact of quantum corrections.}

\section{Phonon spectra and semiclassical limit}
\label{Sec:3}

\begin{figure}
    \centering
    \includegraphics[width=0.9\columnwidth]{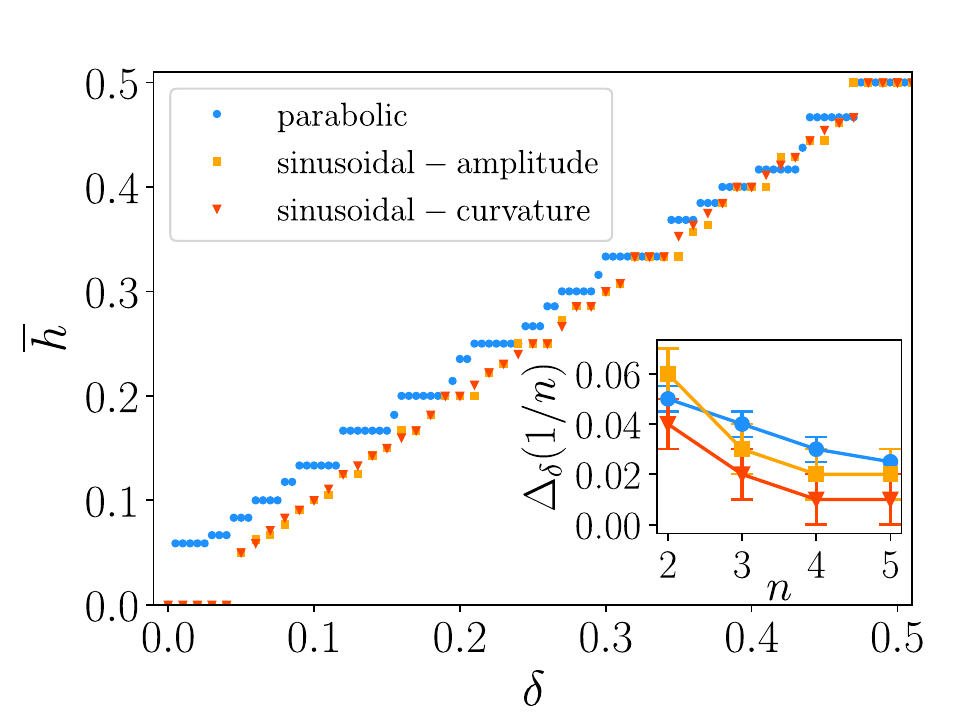}
    \caption{ Magnetization $\overline{h}=m/n$ as a function of the discommensuration $\delta$ for $N=100$ ions. The blue dots refer to the piece-wise parabolic potential with $g=3$, the orange dots to a sinusoidal potential whose depth $V_0$ is the same as the piecewise parabolic potential, the red dots to a sinusoidal potential whose curvature at the minima is the same as the parabolic potential. The staircase for the parabolic potential has been numerically determined using the method of Ref.\ \cite{Koziol:2023}. For the sinusoidal potentials we used a gradient-descent method. The displayed plateaus correspond here to the ratios $m/n$ with $n\leq 20$. The inset compares the width of the plateaus as a function of $n$.}
    \label{fig:11}
\end{figure}

In this section we analyse the low-energy excitations of a Coulomb chain across the transition assuming the temperature is below $T(N)$. In our analysis the spectrum consists of the linear excitations of the classical ground state. We consider an experimentally relevant configuration, where the substrate potential is sinusoidal. Correspondingly, the function in Eq.\ \eqref{Eq:V:0} reads
\begin{equation}
f(x)= 1 - \cos\left( \dfrac{2\pi}{a}x\right) \,.
\end{equation}
This function is continuous and permits us to perform the Taylor expansion about the equilibrium positions for any value of $g$. The corresponding kink, in the continuum limit, is a sine-Gordon soliton with long-range tails \cite{Landa:2020}. Below we refrain to take the continuum limit and keep the discrete nature of the charge distribution.

\begin{figure*}
    \centering
    \includegraphics[width=0.9\textwidth]{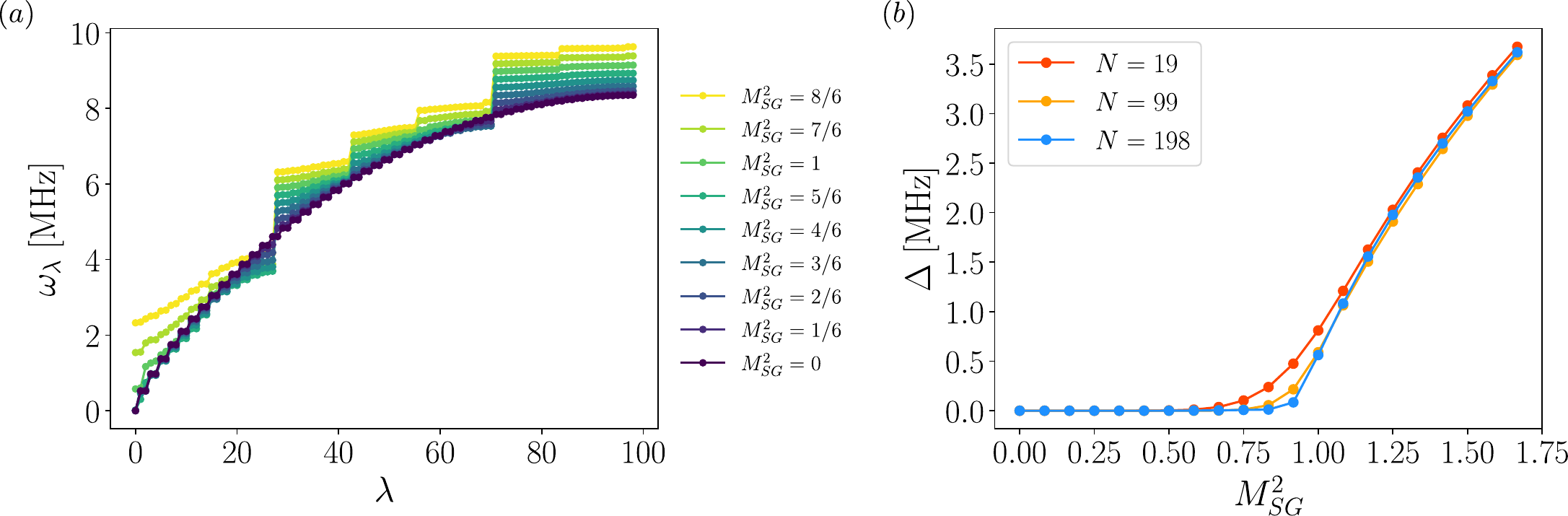}
    \caption{Vibrational spectrum. (a) Eigenfrequencies $\omega_{\lambda}$ for an ion chain and several values of the parameter $M_{SG}$. Here, $\lambda$ labels the eigenmodes for increasing frequency. Here, the density is $\rho = 99/721$ and the particles interact via the Coulomb repulsion ($\alpha = 1$). (b) Spectral gap as a function of $M_{SG}$ for a fixed value of the density $\rho$ and $N=19,99,198$. The calculations are performed using the parameters of \cite{Cetina:2013}, taking $^{174}\mathrm{Yb}^+$ ions with lattice periodicity $a=185\mathrm{nm}$. The interparticle distance $d_0 = 1.35 \mu\mathrm{m}$ is chosen in accordance to $\rho$.}
    \label{fig:7}
\end{figure*}

\begin{figure}
    \centering
    \includegraphics[width=\columnwidth]{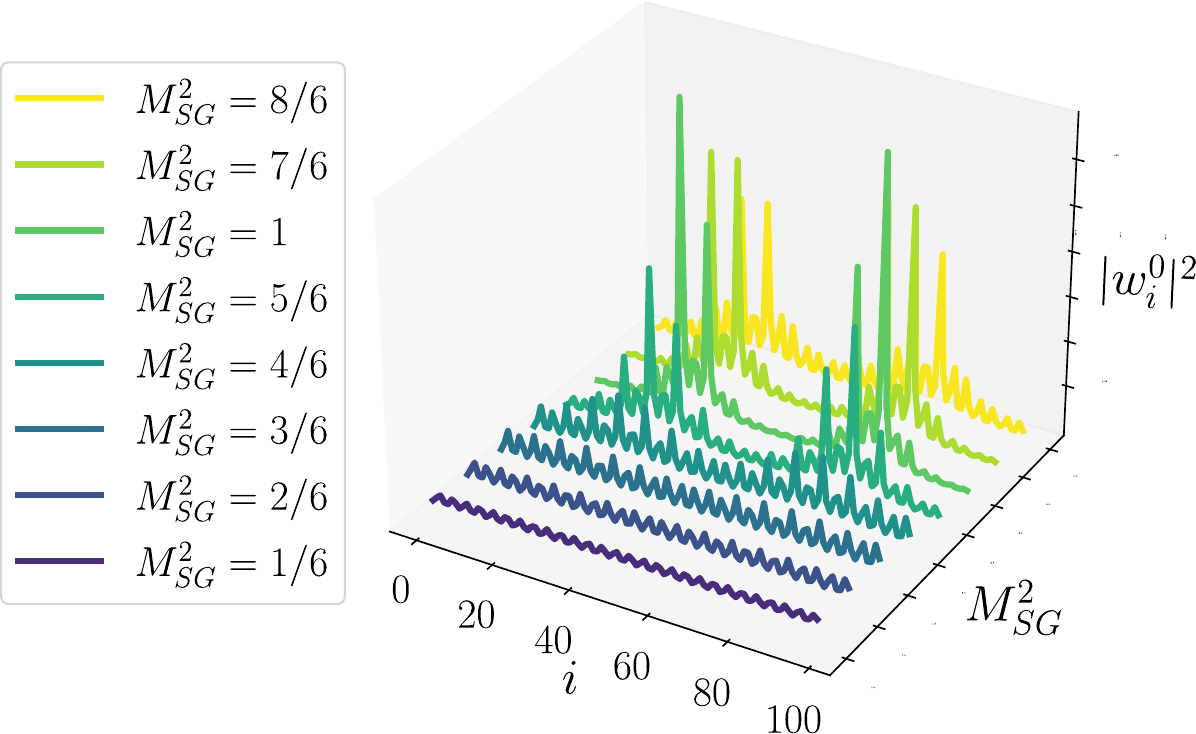}
    \caption{Spatial form of the lowest-energy mode $\vert w^0_i\vert^2$ as a function of the mass of the soliton $M_{SG}$. The considered chain contains $N=99$ ions at a density $\rho=99/721$. Calculations are performed for $^{174}\mathrm{Yb}^{+}$ ions with the characteristic lengths of the experiment described in \cite{Cetina:2013}.}
    \label{fig:8}
\end{figure}
\subsection{Ground state configuration}

The equilibrium positions minimize the potential \eqref{Eq:V:0} and are determined numerically via a gradient-descent algorithm. We denote by $\lbrace \bar x_i^{(0)}\rbrace$ the ensemble of solutions. Defining the local displacement with respect to these equilibrium position as $x_i = \bar x_i^{(0)} + \bar u_i$, the potential energy is expanded up to the second order in the displacement:
\begin{equation}
    H \simeq \dfrac{1}{2M}\sum_i{p^2_i} + \bar V^{(0)} + \dfrac{1}{2}\sum_{i,j}{\bar u_i \mathcal{K}_{ij} \bar u_j} + \mathcal{O}(u^3)\,\label{Eq:Ham},
\end{equation}
where now $\bar V^{(0)}$ is the potential at the equilibrium positions $\lbrace \bar x_i^{(0)}\rbrace$ and 
$\mathcal{K}_{ij}$ are the elements of the symmetric stiffness matrix for the equilibrium configuration of the whole potential. The diagonal elements read
\begin{align}
    \mathcal{K}_{ii}&=
    \left(\dfrac{2\pi}{a}\right)^2 V_0\cos\left(\dfrac{2\pi}{a} \bar x_i^{(0)}\right)+\sum_{k \neq i}{\dfrac{2\alpha(\alpha + 1)W_0}{\vert \bar x_i^{(0)} - \bar x_k^{(0)} \vert^{\alpha + 2}}}\,,
\end{align}
while the off-diagonal elements take the form:    
\begin{align}
    \mathcal{K}_{ij}&=
    -\dfrac{\alpha(\alpha + 1)W_0}{\vert \bar x_i^{(0)} - \bar x_j^{(0)} \vert^{\alpha + 2}}\,.
\end{align}

{The ground state configuration of this ion chain is determined via the minimization of the potential energy. In the regime where $g\ll 1$, the particle are essentially confined to the local minima of the substrate potential: the ground state of the ion chain is described by a devil's staircase similar to the one depicted on Fig.~\ref{fig:5}. Figure~\ref{fig:11} compares the devil's staircases of piece-wise parabolic potential and sinusoidal potentials for $g=3$: the displayed devil's staircases have very similar shapes, despite a shift in the position of the plateaus. Furthermore, based on the prediction of Formula~\eqref{Eq:T}, the comparable widths of the plateaus also suggest that commensurate structures are stable in temperature ranges compatible with the ones deduced for the parabolic case.}

\subsection{Vibrational spectrum}

The vibrational spectrum is found by diagonalizing the quadratic potential with the usual procedure, consisting in identifying the corresponding orthogonal matrix $O\in SO(N)$, such that $O^{-1}=O^T$. In the quadratic form, the Hamiltonian $H$ reads
$$H = V^{(0)} + \sum_\lambda{\left[\frac{1}{2M}p_\lambda^2 + \frac{1}{2}M\omega^2_\lambda u^2_\lambda\right]}\,.$$ 
The eigenvalues given by $M\omega^2_\lambda$ are positive when the equilibrium configuration is stable. The frequencies $\omega_\lambda$ give the dispersion relation, where $\lambda$ labels the eigenmode and is {\it not} the quasi-momentum of the lattice since the potential is generally aperiodic. Figure~\ref{fig:7}(a) displays the vibrational spectrum of the ion chain for different values of $M_{SG}$ across the Aubry transition. An increase of the strength of the substrate potential $V_0$ (and therefore of $M_{SG}$) leads to the opening of gaps in the spectrum. On the other hand, the large-wavelength properties are relatively unperturbed up to a critical value where {the low-frequency spectrum becomes gapped.} 

Figure~\ref{fig:7}(b) shows the value of lowest eigenfrequency, $\Delta = \min_{\lambda}\omega_{\lambda}$, as a function of $M_{SG}$ and for a fixed value of the density. This quantity is an order parameter, that signals the transition between the incommensurate phase, which is self-similar, and the commensurate phase, where the array is pinned by the substrate lattice. {One observes a sudden change from a vanishing gap to a finite one starting from a threshold value of $M^2_{SG}\sim 0.9$. The opening of the gap heralds the transition toward a pinned phase.}

Given that the Hamiltonian is generally not symmetric under discrete translations, the eigenmodes $\lambda$ are not phonon modes. {The dependence of the lowest frequency ones is shown in Fig.\ \ref{fig:8} as a function of the position and the mass of the soliton. This figure shows that, when increasing the substrate potential, and consequently the mass of the soliton, the {site-dependent amplitudes of the lowest-frequency mode depart increasingly from a uniformly distributed form by} displaying several localized excitations. This structure indicates that the mode of lowest frequency does not correspond to a {phonon at} wave vector $k=0$.} 

\subsection{Quantum fluctuations}

The mean-field model is amenable to a semiclassical analysis, which can allow us to estimate its stability against quantum fluctuations. This is done according to this phenomenological ansatz: We quantize the fluctuations about the classical ground state and estimate the maximal size at $T=0$. The commensurate phase is stable when the wave packets of all ions are localized within the corresponding well of the substrate potential. We note that this ansatz is plausible away from the transition point. 

We now spell out the criterion. We denote by $\delta x_i$ the displacement with respect to the closest potential minimum, such that $\vert \delta x_i \vert \leq a/2$. The displacement can be separated into the sum of two contributions: the mean-field displacement $\delta x_i^{(0)}$, that determines the equilibrium position of the ion within the well, and the spatial extension of the wavepacket $\delta u$, which we define as
\begin{equation}
   \delta u=  \max_{i=1,\ldots,N}\sqrt{\vert \delta x_i^2 \vert- \delta x_i^{(0)\,2}}\,.
\end{equation}
We determine $\delta u$ as follows. We first quantize the displacements, $\hat{u}_\lambda = \sqrt{u_{0\lambda}/2}(\hat{a}_\lambda + \hat{a}^\dagger_\lambda)$ and the canonically conjugated momentum $\hat{p}_\lambda = -i\sqrt{(\hbar /2u_{0,\lambda}}(\hat{a}_\lambda - \hat{a}^\dagger_\lambda)$, with $u_{0,\lambda}=\hbar/M\omega_\lambda$ and $[\hat a_\lambda,\hat a^\dagger_{\lambda'}]=\delta_{\lambda,\lambda'}$. The Hamiltonian for the quantum fluctuations is the sum of quantum harmonic oscillators, $H_{q}=\sum_\lambda \hbar \omega_\lambda( \hat{a}^\dagger_\lambda \hat{a}_\lambda + 1/2) $.

The quantum fluctuations can be related to the phonon modes via the relation $u_i = \sum_\lambda w_i^{\lambda} u_\lambda$, where $w_i^{\lambda}$ are the elements of the orthogonal matrix $O$ diagonalizing the quadratic Hamiltonian. Assuming that the system is at temperature $T=0$, then all phonon modes are in their ground state, and $\langle u_i^2\rangle = \sum_\lambda{\frac{\hbar}{2M\omega_\lambda}(w^\lambda_i)^2}$. The total displacement is shown in Fig.~\ref{fig:9} as a function of the mean-field displacement. The dashed line represents the value $|\delta x_i|=a/2$, where the quantum fluctuations become relevant and the mean-field treatment fails. For $M_{SG}>M^{(c)}_{SG}$ the quantum corrections are essentially negligible and the displacement with respect to the local minima of the substrate potential remain below the threshold. As $M_{SG}$ decreases towards the transition value, one can observe an increasing role of the quantum fluctuations. 

This graphic analysis permits us to {roughly} estimate an approximate value  $M_{SG}\simeq 0.97$ at which the semiclassical regime {becomes} invalid. This value lies in the close vicinity of the value $M^{(c)}_{SG}\simeq 0.95$, at which the {classical} transition occurs. The narrow size of the interval where the semiclassical regime breaks down shall be compared with the size of quantum fluctuations, that we identified within a full quantum model \cite{Menu:2023}. There, we showed that quantum effects are scaled by the effective Planck constant
\begin{equation}
   \beta^2 = \left(\dfrac{2\pi}{a}\right)^2 \sqrt{\dfrac{2\hbar^2}{3MK}} 
\end{equation}
Using the parameters $a=185\,\mathrm{nm}$, $M=2.9\times 10^{-25}\,\mathrm{kg}$ and $K=1.9\times 10^{-10}\,\mathrm{N/m}$, this leads to $\beta^2\simeq 0.013$, {which qualitatively agrees with our estimate $\Delta M_{SG}/M_{SG}\sim 0.02$.} 

\subsection{Experimental realization}

\begin{figure}
    \centering
    \includegraphics[width=\columnwidth]{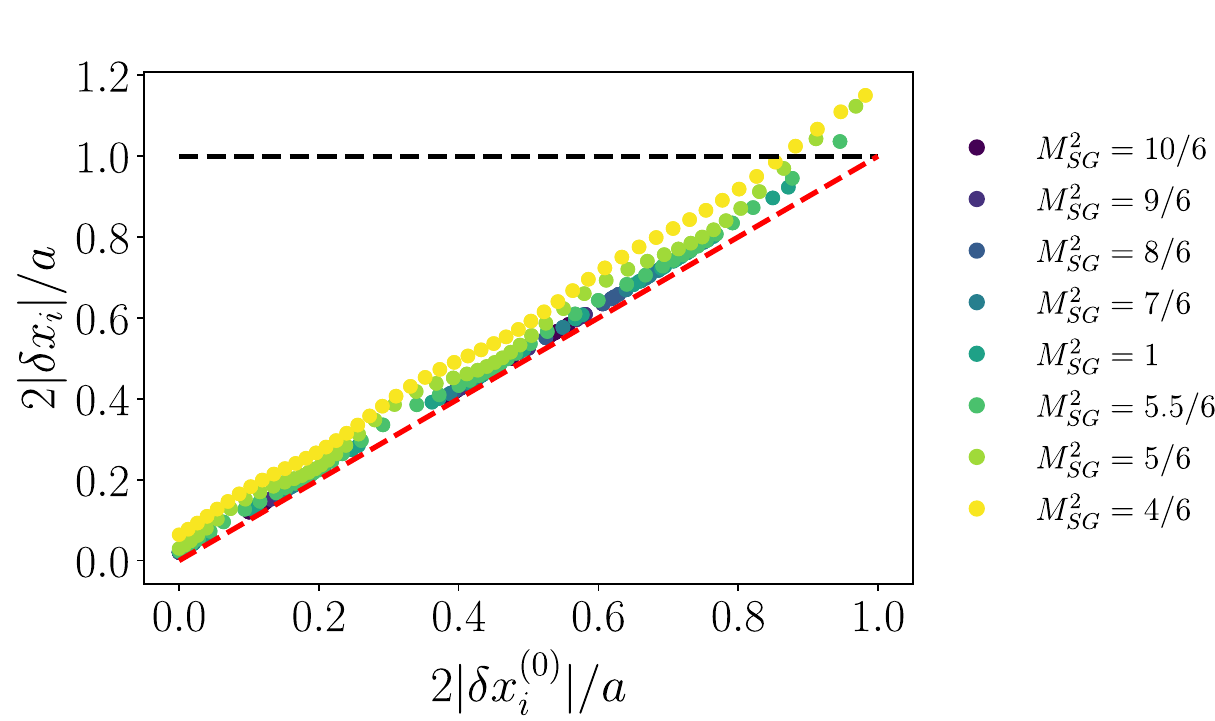}
    \caption{Local displacement of the particles as a function of the mean-field displacement, plotted for a fixed density $\rho = 99/721$ and several values of the soliton mass $M^2_{SG}$. The black dashed line represents the threshold value of $\vert \delta x_i\vert = a/2$, while the red dashed line accounts for the classical behaviour. The parameters are the same as in Fig.~\ref{fig:7}.}
    \label{fig:9}
\end{figure}

The theory developed here is motivated by existing experiments which have measured the Aubry transition in ion chains \cite{Kiethe:2017,Bylinskii:2016}. Their feasibility has been extensively discussed in previous literature \cite{Garcia-Mata,Pruttivarasin:2011,Cetina:2013,Cormick:2013}. In this section we focus on the core assumptions of our work. Throughout this work we have assumed that -in the absence of the optical lattice- the ions are equidistant. In a linear Paul trap, this is fulfilled at the chain center \cite{Morigi:2004,Kiethe:2018}. It is also possible to shape the macrocopic trap potential to approximate a box potential using several electrodes, resulting in a near-homogeneous ion spacing over an extended region. An interesting alternative is offered by ring trap geometries \cite{Haeffner}. Here, a periodic substrate along the ring could be created by a second ion species with different mass \cite{Landa:2014,Fogarty:2013}. Observing reasonably sharp transition requires chains with several tens of ions. Linear chains with 40 ions have been demonstrated \cite{Feng:2020} and laser cooled, and work is under way to extend this number to $\geq 50$.

Kinks and dislocations can be imaged \cite{Pyka, Ulm, Mielenz} and spectroscopically resolved \cite{Brox,Kiethe:2018}. This permits to determine the behavior at the Aubry transition as well as at the commensurate-incommensurate transition. Features of the devil's staircase are visible as long as thermal excitations are smaller than the gap \cite{Shimshoni:2011,Kiethe:2021}. Our study permits to identify the temperatures required: Using the parameters of Ref. \cite{Bylinskii:2016}, for a chain of 100 ions $^{174}$Yb$^+$ with interparticle distance $d_0 = 6 \mu\mathrm{m}$ and lattice periodicity $a=185\mathrm{nm}$, steps of the devil's staircase with magnetization $k=m/n$ will be measurable for temperatures $T\lesssim 1 \,\mathrm{mK}/n$. These temperatures are easily achieved with sideband cooling or EIT cooling \cite{Eschner:2003,Lechner:2016,Feng:2020}, which can reach the quantum ground state of the optical potential, corresponding to temperatures of a few microkelvin.

Quantum effects at the transition manifest as tunnelling of the solitons, and tend to stabilize the commensurate phase. Within our mean-field approach, we have included them as a perturbation and have analysed the corresponding qualitative features numerically. Other studies followed a different ansatz where the soliton tunnels across the Peierls-Nabarro potential \cite{Timm:2021}. The full quantum dynamics has been numerically studied for few ions in \cite{Bonetti}. Finally, in a recent work we derived a mapping valid deep in the incommensurate phase \cite{Menu:2023}. All these considerations lead us to predict that quantum effects at the Aubry and at the commensurate-incommensurate transition should be experimentally observable for ion chain cooled to temperatures $T \lesssim 1$~mK. 

\section{Conclusions}
\label{Sec:5}

We have determined the classical ground state of a Frenkel-Kontorova model with long-range interactions. When the substrate potential is given  by piecewise harmonic oscillators, the long-range Frenkel-Kontorova model can be exactly mapped onto a chain of spin with long-range antiferromagnetic interactions and an external magnetic field. The structure of the coefficients allows us to shed light on the behavior at the commensurate-incommensurate transition and at the Aubry transition. While for power-law interactions scaling as $1/r^\alpha$ and $\alpha>1$ the transitions are well defined also in the thermodynamic limit, for Coulomb interactions, $\alpha=1$, they become a crossover. We have discussed the features signalling the onset of the transitions in an experiment with trapped ions. Importantly, we predict that this transition can still be observed in realistic finite-size experiments, given our analysis of the devil's staircase as a function of the temperature and of the number of ions.

In terms of the theoretical model, our study is complementary to existing works and approaches \cite{Timm:2021,Bonetti,Menu:2023}. The mapping, in fact, allows us to take into full account the discrete nature of the lattice and to assess its role on transitions that are typically characterized in the continuum limit.  The mapping to the model by Ref. \cite{Bak-PRL:1982}, moreover, opens interesting perspectives for studying topological features, characteristic of the fractional quantum Hall effects in ion chains \cite{Rotondo_2016,Nussinov:2016}. 

Finally, our predictions have been derived for a generic power-law exponent, and are in principle applicable to other systems, such as chains of Rydberg atoms in tweezers arrays \cite{Barredo:2016, Endre:2016} or dipoles tightly bound in optically lattices \cite{Lahaye:2009, Baranov:2008}. While a Luttinger liquid description in these cases is more appropriate \cite{Dalmonte:2010}, yet our approach shall allow one to capture finite size effects and the role of discreteness in these dynamics. 
 
Our study contributes to clarify the role of long-range, non-additive interactions on the stability of structures that are commensurate with the external substrate and to identify the regime of stability as a function of the physical parameters. Moreover, it sets a semi-analytic benchmark for numerical investigations of geometric frustration in Coulomb systems. Future studies will focus on  {the analysis of quantum correlations between kinks in these systems, based on the study of Ref.\ \cite{Kahan:2024} for few ions. We will consider} the effect of deformable substrate potentials, as realised in Refs.~\cite{Kiethe:2017,Kiethe:2018} with two ion chains in a linear Paul trap and in Ref.~\cite{Laupretre:2019} by trapping ions in the optical lattice of a high-finesse cavity (see  Ref. \cite{Cormick:2013,Fogarty:2015} for the theoretical predictions in the strong-coupling limit). 

Our results support the present atom-based quantum technology platforms as versatile laboratories to probe condensed-matter and high-energy physics hypothesis.

\section*{Acknowledgments}
The authors acknowledges helpful discussions with and comments of M.-C. Banuls, C. Cormick, E. Demler, S. B. J\"ager, H. Landa, and V. Stojanovic, as well as the contribution of A. A. Buchheit in the preliminary phase of this project. G.M.\ is deeply indebted to C. Bassi-Angelini and E. Auerbach for inspiring comments. R.M. thanks B. Pascal for her precious insight. This work has been supported by the Deutsche Forschungsgemeinschaft (DFG, German Research Foundation), with the CRC-TRR 306 "QuCoLiMa", Project-ID No. 429529648, and by the German Ministry of Education and Research (BMBF) via the project NiQ ("Noise in Quantum Algorithm") and via the QuantERA project NAQUAS. Project NAQUAS has received funding from the QuantERA ERA-NET Cofund in Quantum Technologies implemented within the European Union's Horizon 2020 program. J.Y.M. was supported by the European Social Fund REACT EU through the Italian national program PON 2014-2020, DM MUR 1062/2021. M.L.C. acknowledges support from the National Centre on HPC, Big Data and Quantum Computing - SPOKE 10 (Quantum Computing) and received funding from the European Union Next-GenerationEU - National Recovery and Resilience Plan (NRRP) – MISSION 4 COMPONENT 2, INVESTMENT N. 1.4 – CUP N. I53C22000690001. M.L.C. acknowledge support from the project PRA\_2022\_2023\_98 "IMAGINATION" and acknowledges support from the MIT-UNIPI program. V.V. acknowledges support from the NSF Physics Frontiers Center (PHY-2317134
) and NSF grant PHY-2207996. This research was supported in part by grants NSF PHY-1748958 and PHY-2309135 to the Kavli Institute for Theoretical Physics (KITP).

\begin{appendix}

\section{Determination of the coefficients $J_\alpha(r)$}
\label{App:A}

We here analyse the behavior of the coefficient $J_\alpha(r)$ with the distance using the continuum limit of the sum on the right-hand side of Eq. \eqref{eq:J:0} and performing an analytic continuation. Note that the integral shares several analogies with the integrals performed in Refs.\ \cite{Defenu:2019,Jaeger:2020} for chains with power-law decaying interactions, and arguments applied in those works can be also applied to this case. For convenience, we first define the dimensionless function $B_\alpha(r)=J_\alpha(r)/(4V_0)$. In the continuum limit, it is an integral over the Brillouin zone:
\begin{eqnarray}
B_\alpha(r)\approx\lim_{\eta\to 0^+}\frac{1}{2\pi}\int_\eta^{2\pi-\eta}{\rm d}q \,{\rm e}^{{\rm i}qra}{\mathcal M}_\alpha(q)
\label{eq:Bm}
\end{eqnarray}
with
\begin{equation}
\label{M:q}
{\mathcal M}_\alpha(q)=
\frac{1}{2{\mathcal F}_\alpha(q)}\frac{\phi_\alpha(q)}{\phi_\infty(q)}\,.
\end{equation}
where we have introduced the function
\begin{equation}
\label{F:q}
{\mathcal F}_\alpha(q)=1/g+\phi_\alpha(q)\,.
\end{equation}
It is useful to rewrite the function $\phi_\alpha(q)$, Eq.\ \eqref{Eq:Phi} as
\begin{equation}
\label{Phi:q}
\phi_\alpha(q)=2\zeta(\alpha+2)-{\rm Li}_{\alpha+2}({\rm e}^{{\rm i}qa})-{\rm Li}_{\alpha+2}({\rm e}^{-{\rm i}qa})\,,
\end{equation}
and ${\rm Li}_\gamma(z)=\sum_{n=1}^\infty z^n/n^\gamma$ the polylogarithm \cite{AbramowitzStegun}, while $\zeta(\gamma)=\sum_{n=1}^{\infty} 1/n^\gamma=\mathrm{Li}_\gamma(1)$ stands for the Riemann $\zeta$-function. We also note that in the limit $q\to 0$ the function $\lim_{q\to 0} {\mathcal M}_\alpha(q)=g\zeta(\alpha)/2$ for all values of $\alpha\ge 1$. 

\begin{figure}
	\center \includegraphics[width=0.9\linewidth]{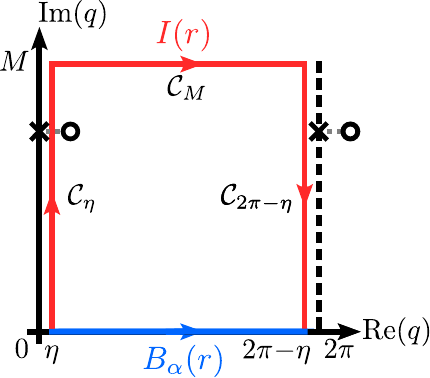}
	\caption{Sketch of the contour integration in \eqref{In}. Here, the blue line indicates the integration path $\mathcal{C}(y)=y$ along the real axis ($y\in[\eta,2\pi-\eta]$). The red line shows the contour $\mathcal{C}_{\eta}(y)=\eta +iy$ with $y\in[0,M]$, $\mathcal{C}_{M}(y)=y+iM$ with $y\in[\eta,2\pi-\eta]$, and $\mathcal{C}_{2\pi-\eta}(y)=2\pi-\eta+i(M-y)$ with $y\in[0,M]$. Black crosses mark the positions of the displaced poles (black circles) in the limit $\epsilon\to 0+$.}
 \label{Fig:10}
\end{figure}

We identify the contour in the complex plane illustrated on Fig.~\ref{Fig:10}. {It consists in a contribution along a segment (in blue) of the real axis, integrating between $[\eta, 2\pi-\eta]$. It accounts for the quantity $B_\alpha(r)$ defined in Eq.~\eqref{eq:Bm}. The path in red is segmented into three contributions: $\mathcal{C}_\eta$ obtained by integrating along the path $z=\eta+iy$, with $y\in [0,M]$, $\mathcal{C}_M$ obtained by integrating along the path $z=y+iM$, with $y\in [\eta,2\pi - \eta]$, and $\mathcal{C}_{2\pi-\eta}$ obtained by integrating along the path $z=2\pi -\eta+i(M-y)$, with $y\in [0,M]$. Summing these three integrals yields the contribution $I(r)$. We also notice that in the limit $\eta\to 0^+$ the contour intersects with poles located in $z=iq_0$ and $z=iq_0 + 2\pi$. This pathological case can however be solved by introducing a shift of the poles by a factor $\epsilon$. As a result, we need to include contributions coming from the pole contained within the contour, namely $z=iq_0$.}

Using the residue theorem, we rewrite Eq.\ \eqref{eq:Bm} as the sum of the integral along the contour, $I(r)$, and of the residues it contains, $R(r)$, as it follows
\begin{align}
B_\alpha(r)= R(r)+I(r)\,.
\label{Cnequ}
\end{align}
The integral is taken along the contour illustrated in Fig.~\ref{Fig:10} and reads
\begin{align}
I(r)=\lim_{\eta\rightarrow 0^+}\left[\int_{\mathcal{C}_\eta}+ \int_{\mathcal{C}_{2\pi-\eta}}+ \int_{\mathcal{C}_M}\right] dq\,\frac{e^{i|r|q}}{2\pi}{\mathcal M}_\alpha(q)\,,\label{In}
\end{align}

The summation over the residues in Eq.~\eqref{Cnequ} goes over the complex numbers $q_0$ with ${0\leq\mathrm{Re}(q_0)<2\pi}$ and ${\mathrm{Im}(r_0)>0}$ for which $\mathrm{Res}[e^{rq}{\mathcal M}(q),q_0]$ does not vanish and reads
\begin{align}
R(r)=i\mathrm{Res}\big[e^{i|r|q}{\mathcal M}_\alpha(q),iq_0\big]
\end{align}
Below we extend the treatment of Ref.\ \cite{Jaeger:2020} to this case and argue that the residues contribute to $B_\alpha(r)$ with an exponential decay (see Eq.~\eqref{R:exp}), while the integral term is different from zero for $\alpha$ finite. In this case its contribution is an algebraic decay with $\alpha$ (see Eq.~\eqref{algebraic}). In particular, we show that it also holds for $\alpha=1$.

\subsection{Exponential decay}
\label{App:A1}
For $\alpha>1$ we observe that the only values where $\mathrm{Res}[e^{irq}{\mathcal M}_\alpha(q),s]$ does not vanish are the zeros of $\mathcal{F}_\alpha(q)$, defined in Eq.~\eqref{F:q}. Therefore we search for the values $s$ such that  $\mathcal{F}_\alpha(s)=0$. Within the area of the contour ${\mathcal M}(q)$ is a meromorphic function and we can take its Laurent series about the root $s$ of $\mathcal{F}_\alpha(q)$:
\begin{align}
{\mathcal M}_\alpha(q)=\sum_{\ell=-\infty}^{\infty}a_{\ell}(s)(q-s)^{\ell}\,,\label{Laurent}
\end{align}
where the $a_{\ell}$ are the coefficients. Moreover, since  ${\mathcal M}_\alpha(q)$ is meromorphic there exists a finite and positive index $L_{s}$ such that $a_{\ell}(s)=0$ for $\ell<-L_{s}$. This index determines  the order of the pole of ${\mathcal M}_\alpha(q)$ at $s$.

Using Eq.~\eqref{Laurent} the residues at $s$ can be expressed as 
\begin{align}
\label{Rn}
\mathrm{Res}\big[e^{i|r|q}{\mathcal M}_\alpha(q),q_0\big]=e^{i|r|s}P_{s}(r)\,,
\end{align}
where $P_{s}(r)$ is a polynomial in $r$ which depends on the coefficients of the Laurent expansion as
\begin{align}
P_{s}(r)=\sum_{\ell=0}^{L_{s}-1}\frac{[i|r|]^{\ell}}{\ell!}a_{-1-\ell}(s),
\end{align}
using the expansion of $e^{i|r|(q-s)}$ close to $s$. 
Since all poles are isolated, then the behavior of Eq.~\eqref{Rn} in the bulk is dominated by the residue at the point $s'$ with the smallest imaginary part, namely ${\rm Im}(s')=\xi$ is such that 
${\xi\le {\rm Im}(s)}$ for all $s$. We distinguish two cases, when $\xi>0$ and when instead $\xi=0$. For $\xi>0$  then for $r\gg 1$ the sum over the residues Eq.~\eqref{Rn} behaves as
\begin{align}
\label{R:exp}
R(r) \sim e^{-\xi |r|}\tilde P_{s'}(r)\,,
\end{align}
%
where $\tilde P_{s'}(r)= iP_{s'}(r)e^{i|r|{\rm Re}(s')}$.  If $s'$ is a pole of order one, then the polynomial $P_{s'}(r)$ is simply a constant independent on $r$.

The residues are simply found for the case $\alpha\to\infty$, where the interaction is nearest-neighbour. Then $\mathcal{F}_\infty(q)$ has two poles, $q_\pm=\pm i\ln(\epsilon+\sqrt{\epsilon^2-1})=\pm iq_0$, where for convenience we have introduced $\epsilon=1+1/(2g)$. Only the pole $q_+$ is within the contour, and we obtain 
\begin{equation}
R(r)=\frac{V_0}{2}\,\left(g\frac{{\rm e}^{-q_0|r|}}{\sqrt{4g+1}}\right)\,.
\end{equation}
This expression agrees with Equation (3.23) of Ref. \cite{Pietronero} (note that their coefficient $g$ is our $g$ divided by 2).

When $\xi=0$, a pole of $F_\alpha(q)$ lies on the real axis. We note that it can occur only for $g\to\infty$, which is outside of the validity of our model.
For $\alpha=1$, instead, there is no simple pole.

\subsection{Power-law tails}
\label{App:A2}
We will now extract the behavior of the integrals in Eq.~\eqref{In}. For this purpose we use that $\int_{\mathcal{C}_M}dqe^{i|r|q}{\mathcal M_\alpha}(q)$ vanishes in the limit $M\to\infty$. In this limit 
the integral to solve is 
\begin{align}
I(r)&=-\frac{1}{\pi}\int_{0}^{+\infty} e^{-y|r|}{\rm Im}({\mathcal M}_\alpha(iy))\mathrm{d}y\,.\label{Integral}
\end{align} 
Here, ${\mathcal M}_\alpha(q)$ is given in Eq.~\eqref{M:q}, and its imaginary part specifically reads:
\begin{eqnarray}
\label{Mk:Im}
& &\mathrm{Im}({\mathcal M}_\alpha(iy))=-\frac{1}{g}\frac{{\rm Im}(\mathcal{F}_\alpha(iy))}{|\mathcal{F}_\alpha(iy)|^2}\,{\rm Re}\left(\frac{1}{2-e^{y}-e^{-y}}\right)\nonumber\\
& &+{\rm Im}\left(\frac{1}{2-e^{y}-e^{-y}}\right)\left(1-\frac{1}{g}\frac{{\rm Re}(\mathcal{F}_\alpha(iy))}{|\mathcal{F}_\alpha(iy)|^2}\right)\,.
\end{eqnarray}
with the function $\mathcal{F}_\alpha(q)$ given by Eq.\ \eqref{F:q}.

In order to determine the behavior for $r\gg 1$, we expand $\mathrm{Im}({\mathcal M}(iy))$ in leading order of $y$ using the Taylor expansion of the Polylogarithm \cite{Olver:2010}:
\begin{align*}
\mathrm{Li}_\gamma(e^{-y})=&\Gamma(1-\gamma)y^{\gamma-1}+\sum_{k=0}^\infty\frac{\zeta(\gamma-k)}{k!}(-y)^k,\\
\mathrm{Li}_\gamma(e^{y})=&\Gamma(1-\gamma)\cos(\pi(\gamma-1))y^{\gamma-1}+\sum_{k=0}^\infty\frac{\zeta(\gamma-k)}{k!}y^k\nonumber\\&+i\Gamma(1-\gamma)\sin(\pi(\gamma-1))y^{\gamma-1}\, .
\end{align*}
Here, the real part is well-defined only for $\gamma\notin \mathbb{N}$, while the coefficient of the imaginary part is ${\Gamma(1-\gamma)\sin(\pi(1-\gamma))}={\pi/\Gamma(\gamma)}$. 
In leading order in the expansion, Eq. \eqref{Mk:Im} is given by 
\begin{align*}
\mathrm{Im}({\mathcal M}_\alpha(iy))\approx-\frac{g}{2}\frac{\pi}{\Gamma(\alpha+2)} y^{\alpha-1}\,.
\end{align*}
Substituting in Eq. \eqref{Integral} we obtain
\begin{align}
I(r) &\approx \frac{g}{2}\dfrac{1}{\Gamma(\alpha+2)}\int_0^{+\infty}{e^{-y \vert r\vert}y^{\alpha - 1}\mathrm{d}y}\nonumber\\
&\approx\frac{g}{2}\,\frac{\Gamma(\alpha)}{\Gamma(\alpha+2)}|r|^{-\alpha}\,,\label{algebraic}
\end{align}
which is valid for $|r|\gg 1$. This expression shows that the integral vanishes for $\alpha\to\infty$, thus in the nearest neighbour case. In this case the coefficient decays as an exponential function. For $1\le\alpha<\infty$, instead, the decay is algebraic with the same exponent as the interaction potential.
For the case we consider here, where $d_0\gg a$ and thus $|r|\gg a$, the algebraic decay determines the coefficients behaviour. Therefore, $B_\alpha(r)$ takes the form given in Eq. \eqref{Eq:J}.

\section{Determination of the displacements $\delta x_j$}
\label{App:B}
The displacements $\delta x_j$ can be expressed in terms of the segments $h_j$ by integrating Eq.~\eqref{Eq:Qq:1}:
\begin{equation}
\delta x_j=a\sum_r{\mathcal F}(r)(h_{j+r}-\langle h\rangle)\,,
\end{equation}
where
\begin{eqnarray}
{\mathcal F}(r)=\frac{1}{N}\sum_q\frac{{\rm e}^{iqra}}{1-{\rm e}^{-iqa}}\left(1-\frac{1/g}{\mathcal{F}_\alpha(q)}\right)\,\label{F:r},
\end{eqnarray}
and $\mathcal{F}_\alpha(q)$ is given in Eq.\ \eqref{F:q}
Here, ${\rm Li}_\alpha(y)=\sum_{n=1}^\infty y^n/n^\alpha$ is the Polylogarithm, $|y|\le 1$, and $\zeta(\alpha)={\rm Li}_\alpha(1)$ is Riemann's zeta function \cite{Olver:2010}.
Using Eq. \eqref{Eq:Qq:1} we can write the displacements $\delta x_j$ defined in Eq. \eqref{Eq:h:0}, as a function of the configuration of empty sequences. Using the analytic continuation  as shown above we find an explicit expression for $\mathcal F(r)$ and thus for the displacements from the well centres as a function of the empty sequences: 
\begin{eqnarray}
\delta x_j &\approx & a\frac{g}{\alpha+1}\sum_{r>0}\frac{1}{r^{\alpha+1}}(h_{j+r}-h_{j-r})\,.
\label{eq:delta:x}
\end{eqnarray}
This expression shows that the displacement counterbalances the net force due to the surrounding ions.

\end{appendix}
\bibliography{biblio}

\begin{thebibliography}{77}%
\makeatletter
\providecommand \@ifxundefined [1]{%
 \@ifx{#1\undefined}
}%
\providecommand \@ifnum [1]{%
 \ifnum #1\expandafter \@firstoftwo
 \else \expandafter \@secondoftwo
 \fi
}%
\providecommand \@ifx [1]{%
 \ifx #1\expandafter \@firstoftwo
 \else \expandafter \@secondoftwo
 \fi
}%
\providecommand \natexlab [1]{#1}%
\providecommand \enquote  [1]{``#1''}%
\providecommand \bibnamefont  [1]{#1}%
\providecommand \bibfnamefont [1]{#1}%
\providecommand \citenamefont [1]{#1}%
\providecommand \href@noop [0]{\@secondoftwo}%
\providecommand \href [0]{\begingroup \@sanitize@url \@href}%
\providecommand \@href[1]{\@@startlink{#1}\@@href}%
\providecommand \@@href[1]{\endgroup#1\@@endlink}%
\providecommand \@sanitize@url [0]{\catcode `\\12\catcode `\$12\catcode
  `\&12\catcode `\#12\catcode `\^12\catcode `\_12\catcode `\%12\relax}%
\providecommand \@@startlink[1]{}%
\providecommand \@@endlink[0]{}%
\providecommand \url  [0]{\begingroup\@sanitize@url \@url }%
\providecommand \@url [1]{\endgroup\@href {#1}{\urlprefix }}%
\providecommand \urlprefix  [0]{URL }%
\providecommand \Eprint [0]{\href }%
\providecommand \doibase [0]{http://dx.doi.org/}%
\providecommand \selectlanguage [0]{\@gobble}%
\providecommand \bibinfo  [0]{\@secondoftwo}%
\providecommand \bibfield  [0]{\@secondoftwo}%
\providecommand \translation [1]{[#1]}%
\providecommand \BibitemOpen [0]{}%
\providecommand \bibitemStop [0]{}%
\providecommand \bibitemNoStop [0]{.\EOS\space}%
\providecommand \EOS [0]{\spacefactor3000\relax}%
\providecommand \BibitemShut  [1]{\csname bibitem#1\endcsname}%
\let\auto@bib@innerbib\@empty
\bibitem [{\citenamefont {Dubin}\ and\ \citenamefont
  {O'Neil}(1999)}]{Dubin:RMP}%
  \BibitemOpen
  \bibfield  {author} {\bibinfo {author} {\bibfnamefont {D.~H.~E.}\
  \bibnamefont {Dubin}}\ and\ \bibinfo {author} {\bibfnamefont {T.~M.}\
  \bibnamefont {O'Neil}},\ }\bibfield  {title} {\enquote {\bibinfo {title}
  {Trapped nonneutral plasmas, liquids, and crystals (the thermal equilibrium
  states)},}\ }\href {\doibase 10.1103/RevModPhys.71.87} {\bibfield  {journal}
  {\bibinfo  {journal} {Rev. Mod. Phys.}\ }\textbf {\bibinfo {volume} {71}},\
  \bibinfo {pages} {87--172} (\bibinfo {year} {1999})}\BibitemShut {NoStop}%
\bibitem [{\citenamefont {Schulz}(1993)}]{Schulz:1993}%
  \BibitemOpen
  \bibfield  {author} {\bibinfo {author} {\bibfnamefont {H.~J.}\ \bibnamefont
  {Schulz}},\ }\bibfield  {title} {\enquote {\bibinfo {title} {Wigner crystal
  in one dimension},}\ }\href {\doibase 10.1103/PhysRevLett.71.1864} {\bibfield
   {journal} {\bibinfo  {journal} {Phys. Rev. Lett.}\ }\textbf {\bibinfo
  {volume} {71}},\ \bibinfo {pages} {1864--1867} (\bibinfo {year}
  {1993})}\BibitemShut {NoStop}%
\bibitem [{\citenamefont {Morigi}\ and\ \citenamefont
  {Fishman}(2004)}]{Morigi:2004}%
  \BibitemOpen
  \bibfield  {author} {\bibinfo {author} {\bibfnamefont {G.}~\bibnamefont
  {Morigi}}\ and\ \bibinfo {author} {\bibfnamefont {S.}~\bibnamefont
  {Fishman}},\ }\bibfield  {title} {\enquote {\bibinfo {title} {Dynamics of an
  ion chain in a harmonic potential},}\ }\href {\doibase
  10.1103/PhysRevE.70.066141} {\bibfield  {journal} {\bibinfo  {journal} {Phys.
  Rev. E}\ }\textbf {\bibinfo {volume} {70}},\ \bibinfo {pages} {066141}
  (\bibinfo {year} {2004})}\BibitemShut {NoStop}%
\bibitem [{\citenamefont {Birkl}\ \emph {et~al.}(1992)\citenamefont {Birkl},
  \citenamefont {Kassner},\ and\ \citenamefont {Walther}}]{Birkl:1992}%
  \BibitemOpen
  \bibfield  {author} {\bibinfo {author} {\bibfnamefont {G.}~\bibnamefont
  {Birkl}}, \bibinfo {author} {\bibfnamefont {S.}~\bibnamefont {Kassner}}, \
  and\ \bibinfo {author} {\bibfnamefont {H.}~\bibnamefont {Walther}},\
  }\bibfield  {title} {\enquote {\bibinfo {title} {{Multiple-shell structures
  of laser-cooled $^{24} \mathrm{Mg}^+$ ions in a quadrupole storage ring}},}\
  }\href {\doibase 10.1038/357310a0} {\bibfield  {journal} {\bibinfo  {journal}
  {Nature}\ }\textbf {\bibinfo {volume} {357}},\ \bibinfo {pages} {310--313}
  (\bibinfo {year} {1992})}\BibitemShut {NoStop}%
\bibitem [{\citenamefont {Raizen}\ \emph {et~al.}(1992)\citenamefont {Raizen},
  \citenamefont {Gilligan}, \citenamefont {Bergquist}, \citenamefont {Itano},\
  and\ \citenamefont {Wineland}}]{Raizen:1992}%
  \BibitemOpen
  \bibfield  {author} {\bibinfo {author} {\bibfnamefont {M.~G.}\ \bibnamefont
  {Raizen}}, \bibinfo {author} {\bibfnamefont {J.~M.}\ \bibnamefont
  {Gilligan}}, \bibinfo {author} {\bibfnamefont {J.~C.}\ \bibnamefont
  {Bergquist}}, \bibinfo {author} {\bibfnamefont {W.~M.}\ \bibnamefont
  {Itano}}, \ and\ \bibinfo {author} {\bibfnamefont {D.~J.}\ \bibnamefont
  {Wineland}},\ }\bibfield  {title} {\enquote {\bibinfo {title} {{Ionic
  crystals in a linear Paul trap}},}\ }\href {\doibase
  10.1103/PhysRevA.45.6493} {\bibfield  {journal} {\bibinfo  {journal} {Phys.
  Rev. A}\ }\textbf {\bibinfo {volume} {45}},\ \bibinfo {pages} {6493--6501}
  (\bibinfo {year} {1992})}\BibitemShut {NoStop}%
\bibitem [{\citenamefont {Dubin}(1993)}]{Dubin:1993}%
  \BibitemOpen
  \bibfield  {author} {\bibinfo {author} {\bibfnamefont {D.~H.~E.}\
  \bibnamefont {Dubin}},\ }\bibfield  {title} {\enquote {\bibinfo {title}
  {Theory of structural phase transitions in a trapped coulomb crystal},}\
  }\href {\doibase 10.1103/PhysRevLett.71.2753} {\bibfield  {journal} {\bibinfo
   {journal} {Phys. Rev. Lett.}\ }\textbf {\bibinfo {volume} {71}},\ \bibinfo
  {pages} {2753--2756} (\bibinfo {year} {1993})}\BibitemShut {NoStop}%
\bibitem [{\citenamefont {Fishman}\ \emph {et~al.}(2008)\citenamefont
  {Fishman}, \citenamefont {De~Chiara}, \citenamefont {Calarco},\ and\
  \citenamefont {Morigi}}]{Fishman:2008}%
  \BibitemOpen
  \bibfield  {author} {\bibinfo {author} {\bibfnamefont {S.}~\bibnamefont
  {Fishman}}, \bibinfo {author} {\bibfnamefont {G.}~\bibnamefont {De~Chiara}},
  \bibinfo {author} {\bibfnamefont {T.}~\bibnamefont {Calarco}}, \ and\
  \bibinfo {author} {\bibfnamefont {G.}~\bibnamefont {Morigi}},\ }\bibfield
  {title} {\enquote {\bibinfo {title} {Structural phase transitions in
  low-dimensional ion crystals},}\ }\href {\doibase 10.1103/PhysRevB.77.064111}
  {\bibfield  {journal} {\bibinfo  {journal} {Phys. Rev. B}\ }\textbf {\bibinfo
  {volume} {77}},\ \bibinfo {pages} {064111} (\bibinfo {year}
  {2008})}\BibitemShut {NoStop}%
\bibitem [{\citenamefont {Ulm}\ \emph {et~al.}(2013)\citenamefont {Ulm},
  \citenamefont {Roßnagel}, \citenamefont {Jacob}, \citenamefont {Degünther},
  \citenamefont {Dawkins}, \citenamefont {Poschinger}, \citenamefont
  {Nigmatullin}, \citenamefont {Retzker}, \citenamefont {Plenio}, \citenamefont
  {Schmidt-Kaler},\ and\ \citenamefont {Singer}}]{Ulm}%
  \BibitemOpen
  \bibfield  {author} {\bibinfo {author} {\bibfnamefont {S.}~\bibnamefont
  {Ulm}}, \bibinfo {author} {\bibfnamefont {J.}~\bibnamefont {Roßnagel}},
  \bibinfo {author} {\bibfnamefont {G.}~\bibnamefont {Jacob}}, \bibinfo
  {author} {\bibfnamefont {C.}~\bibnamefont {Degünther}}, \bibinfo {author}
  {\bibfnamefont {S.~T.}\ \bibnamefont {Dawkins}}, \bibinfo {author}
  {\bibfnamefont {U.~G.}\ \bibnamefont {Poschinger}}, \bibinfo {author}
  {\bibfnamefont {R.}~\bibnamefont {Nigmatullin}}, \bibinfo {author}
  {\bibfnamefont {A.}~\bibnamefont {Retzker}}, \bibinfo {author} {\bibfnamefont
  {M.~B.}\ \bibnamefont {Plenio}}, \bibinfo {author} {\bibfnamefont
  {F.}~\bibnamefont {Schmidt-Kaler}}, \ and\ \bibinfo {author} {\bibfnamefont
  {K.}~\bibnamefont {Singer}},\ }\bibfield  {title} {\enquote {\bibinfo {title}
  {Observation of the {Kibble}–{Zurek} scaling law for defect formation in
  ion crystals},}\ }\href {\doibase 10.1038/ncomms3290} {\bibfield  {journal}
  {\bibinfo  {journal} {Nat Commun}\ }\textbf {\bibinfo {volume} {4}},\
  \bibinfo {pages} {2290} (\bibinfo {year} {2013})}\BibitemShut {NoStop}%
\bibitem [{\citenamefont {Pyka}\ \emph {et~al.}(2013)\citenamefont {Pyka},
  \citenamefont {Keller}, \citenamefont {Partner}, \citenamefont {Nigmatullin},
  \citenamefont {Burgermeister}, \citenamefont {Meier}, \citenamefont
  {Kuhlmann}, \citenamefont {Retzker}, \citenamefont {Plenio}, \citenamefont
  {Zurek}, \citenamefont {del Campo},\ and\ \citenamefont
  {Mehlst\"aubler}}]{Pyka}%
  \BibitemOpen
  \bibfield  {author} {\bibinfo {author} {\bibfnamefont {K.}~\bibnamefont
  {Pyka}}, \bibinfo {author} {\bibfnamefont {J.}~\bibnamefont {Keller}},
  \bibinfo {author} {\bibfnamefont {H.L.}\ \bibnamefont {Partner}}, \bibinfo
  {author} {\bibfnamefont {R.}~\bibnamefont {Nigmatullin}}, \bibinfo {author}
  {\bibfnamefont {T.}~\bibnamefont {Burgermeister}}, \bibinfo {author}
  {\bibfnamefont {D.M.}\ \bibnamefont {Meier}}, \bibinfo {author}
  {\bibfnamefont {K.}~\bibnamefont {Kuhlmann}}, \bibinfo {author}
  {\bibfnamefont {A.}~\bibnamefont {Retzker}}, \bibinfo {author} {\bibfnamefont
  {M.B.}\ \bibnamefont {Plenio}}, \bibinfo {author} {\bibfnamefont {W.H.}\
  \bibnamefont {Zurek}}, \bibinfo {author} {\bibfnamefont {A.}~\bibnamefont
  {del Campo}}, \ and\ \bibinfo {author} {\bibfnamefont {T.E.}\ \bibnamefont
  {Mehlst\"aubler}},\ }\bibfield  {title} {\enquote {\bibinfo {title}
  {{Topological defect formation and spontaneous symmetry breaking in ion
  Coulomb crystals}},}\ }\href {https://www.nature.com/articles/ncomms3291}
  {\bibfield  {journal} {\bibinfo  {journal} {Nature Commun.}\ }\textbf
  {\bibinfo {volume} {4}},\ \bibinfo {pages} {2291} (\bibinfo {year}
  {2013})}\BibitemShut {NoStop}%
\bibitem [{\citenamefont {Mielenz}\ \emph {et~al.}(2013)\citenamefont
  {Mielenz}, \citenamefont {Brox}, \citenamefont {Kahra}, \citenamefont
  {Leschhorn}, \citenamefont {Albert}, \citenamefont {Schaetz}, \citenamefont
  {Landa},\ and\ \citenamefont {Reznik}}]{Mielenz}%
  \BibitemOpen
  \bibfield  {author} {\bibinfo {author} {\bibfnamefont {M.}~\bibnamefont
  {Mielenz}}, \bibinfo {author} {\bibfnamefont {J.}~\bibnamefont {Brox}},
  \bibinfo {author} {\bibfnamefont {S.}~\bibnamefont {Kahra}}, \bibinfo
  {author} {\bibfnamefont {G.}~\bibnamefont {Leschhorn}}, \bibinfo {author}
  {\bibfnamefont {M.}~\bibnamefont {Albert}}, \bibinfo {author} {\bibfnamefont
  {T.}~\bibnamefont {Schaetz}}, \bibinfo {author} {\bibfnamefont
  {H.}~\bibnamefont {Landa}}, \ and\ \bibinfo {author} {\bibfnamefont
  {B.}~\bibnamefont {Reznik}},\ }\bibfield  {title} {\enquote {\bibinfo {title}
  {{Trapping of Topological-Structural Defects in Coulomb Crystals}},}\ }\href
  {\doibase 10.1103/PhysRevLett.110.133004} {\bibfield  {journal} {\bibinfo
  {journal} {Phys. Rev. Lett.}\ }\textbf {\bibinfo {volume} {110}},\ \bibinfo
  {pages} {133004} (\bibinfo {year} {2013})}\BibitemShut {NoStop}%
\bibitem [{\citenamefont {Ejtemaee}\ and\ \citenamefont
  {Haljan}(2013)}]{Ejtemaee:2013}%
  \BibitemOpen
  \bibfield  {author} {\bibinfo {author} {\bibfnamefont {S.}~\bibnamefont
  {Ejtemaee}}\ and\ \bibinfo {author} {\bibfnamefont {P.~C.}\ \bibnamefont
  {Haljan}},\ }\bibfield  {title} {\enquote {\bibinfo {title} {Spontaneous
  nucleation and dynamics of kink defects in zigzag arrays of trapped ions},}\
  }\href {\doibase 10.1103/PhysRevA.87.051401} {\bibfield  {journal} {\bibinfo
  {journal} {Phys. Rev. A}\ }\textbf {\bibinfo {volume} {87}},\ \bibinfo
  {pages} {051401} (\bibinfo {year} {2013})}\BibitemShut {NoStop}%
\bibitem [{\citenamefont {Brox}\ \emph
  {et~al.}(2017{\natexlab{a}})\citenamefont {Brox}, \citenamefont {Kiefer},
  \citenamefont {Bujak}, \citenamefont {Schaetz},\ and\ \citenamefont
  {Landa}}]{Brox:2017}%
  \BibitemOpen
  \bibfield  {author} {\bibinfo {author} {\bibfnamefont {J.}~\bibnamefont
  {Brox}}, \bibinfo {author} {\bibfnamefont {P.}~\bibnamefont {Kiefer}},
  \bibinfo {author} {\bibfnamefont {M.}~\bibnamefont {Bujak}}, \bibinfo
  {author} {\bibfnamefont {T.}~\bibnamefont {Schaetz}}, \ and\ \bibinfo
  {author} {\bibfnamefont {H.}~\bibnamefont {Landa}},\ }\bibfield  {title}
  {\enquote {\bibinfo {title} {Spectroscopy and directed transport of
  topological solitons in crystals of trapped ions},}\ }\href {\doibase
  10.1103/PhysRevLett.119.153602} {\bibfield  {journal} {\bibinfo  {journal}
  {Phys. Rev. Lett.}\ }\textbf {\bibinfo {volume} {119}},\ \bibinfo {pages}
  {153602} (\bibinfo {year} {2017}{\natexlab{a}})}\BibitemShut {NoStop}%
\bibitem [{\citenamefont {Kiethe}\ \emph {et~al.}(2017)\citenamefont {Kiethe},
  \citenamefont {Nigmatullin}, \citenamefont {Kalincev}, \citenamefont
  {Schmirander},\ and\ \citenamefont {Mehlst\"aubler}}]{Kiethe:2017}%
  \BibitemOpen
  \bibfield  {author} {\bibinfo {author} {\bibfnamefont {J.}~\bibnamefont
  {Kiethe}}, \bibinfo {author} {\bibfnamefont {R.}~\bibnamefont {Nigmatullin}},
  \bibinfo {author} {\bibfnamefont {D.}~\bibnamefont {Kalincev}}, \bibinfo
  {author} {\bibfnamefont {T.}~\bibnamefont {Schmirander}}, \ and\ \bibinfo
  {author} {\bibfnamefont {T.E.}\ \bibnamefont {Mehlst\"aubler}},\ }\bibfield
  {title} {\enquote {\bibinfo {title} {{Probing nanofriction and Aubry-type
  signatures in a finite self-organized system}},}\ }\href
  {https://www.nature.com/articles/ncomms15364} {\bibfield  {journal} {\bibinfo
   {journal} {Nature Commun.}\ }\textbf {\bibinfo {volume} {8}},\ \bibinfo
  {pages} {15364} (\bibinfo {year} {2017})}\BibitemShut {NoStop}%
\bibitem [{\citenamefont {Kiethe}\ \emph {et~al.}(2018)\citenamefont {Kiethe},
  \citenamefont {Nigmatullin}, \citenamefont {Schmirander}, \citenamefont
  {Kalincev},\ and\ \citenamefont {Mehlst\"aubler}}]{Kiethe:2018}%
  \BibitemOpen
  \bibfield  {author} {\bibinfo {author} {\bibfnamefont {J.}~\bibnamefont
  {Kiethe}}, \bibinfo {author} {\bibfnamefont {R.}~\bibnamefont {Nigmatullin}},
  \bibinfo {author} {\bibfnamefont {T.}~\bibnamefont {Schmirander}}, \bibinfo
  {author} {\bibfnamefont {D.}~\bibnamefont {Kalincev}}, \ and\ \bibinfo
  {author} {\bibfnamefont {T.~E.}\ \bibnamefont {Mehlst\"aubler}},\ }\bibfield
  {title} {\enquote {\bibinfo {title} {{Nanofriction and motion of topological
  defects in self-organized ion Coulomb crystals}},}\ }\href {\doibase
  10.1088/1367-2630/aaf3d5} {\bibfield  {journal} {\bibinfo  {journal} {New
  Journal of Physics}\ }\textbf {\bibinfo {volume} {20}},\ \bibinfo {pages}
  {123017} (\bibinfo {year} {2018})}\BibitemShut {NoStop}%
\bibitem [{\citenamefont {Gangloff}\ \emph {et~al.}(2020)\citenamefont
  {Gangloff}, \citenamefont {Bylinskii},\ and\ \citenamefont
  {Vuleti\'{c}}}]{Gangloff:2022}%
  \BibitemOpen
  \bibfield  {author} {\bibinfo {author} {\bibfnamefont {D.~A.}\ \bibnamefont
  {Gangloff}}, \bibinfo {author} {\bibfnamefont {A.}~\bibnamefont {Bylinskii}},
  \ and\ \bibinfo {author} {\bibfnamefont {V.}~\bibnamefont {Vuleti\'{c}}},\
  }\bibfield  {title} {\enquote {\bibinfo {title} {{Kinks and nanofriction:
  Structural phases in few-atom chains}},}\ }\href {\doibase
  10.1103/PhysRevResearch.2.013380} {\bibfield  {journal} {\bibinfo  {journal}
  {Phys. Rev. Research}\ }\textbf {\bibinfo {volume} {2}},\ \bibinfo {pages}
  {013380} (\bibinfo {year} {2020})}\BibitemShut {NoStop}%
\bibitem [{\citenamefont {Eschner}\ \emph {et~al.}(2003)\citenamefont
  {Eschner}, \citenamefont {Morigi}, \citenamefont {Schmidt-Kaler},\ and\
  \citenamefont {Blatt}}]{Eschner:2003}%
  \BibitemOpen
  \bibfield  {author} {\bibinfo {author} {\bibfnamefont {J.}~\bibnamefont
  {Eschner}}, \bibinfo {author} {\bibfnamefont {G.}~\bibnamefont {Morigi}},
  \bibinfo {author} {\bibfnamefont {F.}~\bibnamefont {Schmidt-Kaler}}, \ and\
  \bibinfo {author} {\bibfnamefont {R.}~\bibnamefont {Blatt}},\ }\bibfield
  {title} {\enquote {\bibinfo {title} {Laser cooling of trapped ions},}\ }\href
  {\doibase 10.1364/JOSAB.20.001003} {\bibfield  {journal} {\bibinfo  {journal}
  {J. Opt. Soc. Am. B}\ }\textbf {\bibinfo {volume} {20}},\ \bibinfo {pages}
  {1003--1015} (\bibinfo {year} {2003})}\BibitemShut {NoStop}%
\bibitem [{\citenamefont {Shimshoni}\ \emph {et~al.}(2011)\citenamefont
  {Shimshoni}, \citenamefont {Morigi},\ and\ \citenamefont
  {Fishman}}]{Shimshoni:2011}%
  \BibitemOpen
  \bibfield  {author} {\bibinfo {author} {\bibfnamefont {E.}~\bibnamefont
  {Shimshoni}}, \bibinfo {author} {\bibfnamefont {G.}~\bibnamefont {Morigi}}, \
  and\ \bibinfo {author} {\bibfnamefont {S.}~\bibnamefont {Fishman}},\
  }\bibfield  {title} {\enquote {\bibinfo {title} {Quantum zigzag transition in
  ion chains},}\ }\href {\doibase 10.1103/PhysRevLett.106.010401} {\bibfield
  {journal} {\bibinfo  {journal} {Phys. Rev. Lett.}\ }\textbf {\bibinfo
  {volume} {106}},\ \bibinfo {pages} {010401} (\bibinfo {year}
  {2011})}\BibitemShut {NoStop}%
\bibitem [{\citenamefont {Zhang}\ \emph {et~al.}(2023)\citenamefont {Zhang},
  \citenamefont {Chow}, \citenamefont {Ejtemaee},\ and\ \citenamefont
  {Haljan}}]{Zhang:2023}%
  \BibitemOpen
  \bibfield  {author} {\bibinfo {author} {\bibfnamefont {J.}~\bibnamefont
  {Zhang}}, \bibinfo {author} {\bibfnamefont {B.~T.}\ \bibnamefont {Chow}},
  \bibinfo {author} {\bibfnamefont {S.}~\bibnamefont {Ejtemaee}}, \ and\
  \bibinfo {author} {\bibfnamefont {P.~C.}\ \bibnamefont {Haljan}},\ }\bibfield
   {title} {\enquote {\bibinfo {title} {{Spectroscopic characterization of the
  quantum linear-zigzag transition in trapped ions}},}\ }\href {\doibase
  10.1038/s41534-023-00741-5} {\bibfield  {journal} {\bibinfo  {journal} {npj
  Quantum Information}\ }\textbf {\bibinfo {volume} {9}},\ \bibinfo {pages}
  {68} (\bibinfo {year} {2023})}\BibitemShut {NoStop}%
\bibitem [{\citenamefont {Bonetti}\ \emph {et~al.}(2021)\citenamefont
  {Bonetti}, \citenamefont {Rucci}, \citenamefont {Chiofalo},\ and\
  \citenamefont {Vuleti\'c}}]{Bonetti}%
  \BibitemOpen
  \bibfield  {author} {\bibinfo {author} {\bibfnamefont {P.~M.}\ \bibnamefont
  {Bonetti}}, \bibinfo {author} {\bibfnamefont {A.}~\bibnamefont {Rucci}},
  \bibinfo {author} {\bibfnamefont {M.~L.}\ \bibnamefont {Chiofalo}}, \ and\
  \bibinfo {author} {\bibfnamefont {V.}~\bibnamefont {Vuleti\'c}},\ }\bibfield
  {title} {\enquote {\bibinfo {title} {{Quantum effects in the Aubry
  transition}},}\ }\href {\doibase 10.1103/PhysRevResearch.3.013031} {\bibfield
   {journal} {\bibinfo  {journal} {Phys. Rev. Res.}\ }\textbf {\bibinfo
  {volume} {3}},\ \bibinfo {pages} {013031} (\bibinfo {year}
  {2021})}\BibitemShut {NoStop}%
\bibitem [{\citenamefont {Timm}\ \emph {et~al.}(2021)\citenamefont {Timm},
  \citenamefont {R\"uffert}, \citenamefont {Weimer}, \citenamefont {Santos},\
  and\ \citenamefont {Mehlst\"aubler}}]{Timm:2021}%
  \BibitemOpen
  \bibfield  {author} {\bibinfo {author} {\bibfnamefont {L.}~\bibnamefont
  {Timm}}, \bibinfo {author} {\bibfnamefont {L.~A.}\ \bibnamefont {R\"uffert}},
  \bibinfo {author} {\bibfnamefont {H.}~\bibnamefont {Weimer}}, \bibinfo
  {author} {\bibfnamefont {L.}~\bibnamefont {Santos}}, \ and\ \bibinfo {author}
  {\bibfnamefont {T.~E.}\ \bibnamefont {Mehlst\"aubler}},\ }\bibfield  {title}
  {\enquote {\bibinfo {title} {Quantum nanofriction in trapped ion chains with
  a topological defect},}\ }\href {\doibase 10.1103/PhysRevResearch.3.043141}
  {\bibfield  {journal} {\bibinfo  {journal} {Phys. Rev. Res.}\ }\textbf
  {\bibinfo {volume} {3}},\ \bibinfo {pages} {043141} (\bibinfo {year}
  {2021})}\BibitemShut {NoStop}%
\bibitem [{\citenamefont {Vanossi}\ \emph {et~al.}(2013)\citenamefont
  {Vanossi}, \citenamefont {Manini}, \citenamefont {Urbakh}, \citenamefont
  {Zapperi},\ and\ \citenamefont {Tosatti}}]{Vanossi:2013}%
  \BibitemOpen
  \bibfield  {author} {\bibinfo {author} {\bibfnamefont {A.}~\bibnamefont
  {Vanossi}}, \bibinfo {author} {\bibfnamefont {N.}~\bibnamefont {Manini}},
  \bibinfo {author} {\bibfnamefont {M.}~\bibnamefont {Urbakh}}, \bibinfo
  {author} {\bibfnamefont {.}~\bibnamefont {Zapperi}}, \ and\ \bibinfo {author}
  {\bibfnamefont {E.}~\bibnamefont {Tosatti}},\ }\bibfield  {title} {\enquote
  {\bibinfo {title} {Colloquium: Modeling friction: From nanoscale to
  mesoscale},}\ }\href {\doibase 10.1103/RevModPhys.85.529} {\bibfield
  {journal} {\bibinfo  {journal} {Rev. Mod. Phys.}\ }\textbf {\bibinfo {volume}
  {85}},\ \bibinfo {pages} {529--552} (\bibinfo {year} {2013})}\BibitemShut
  {NoStop}%
\bibitem [{\citenamefont {Zanca}\ \emph {et~al.}(2018)\citenamefont {Zanca},
  \citenamefont {Pellegrini}, \citenamefont {Santoro},\ and\ \citenamefont
  {Tosatti}}]{TosattiPNAS}%
  \BibitemOpen
  \bibfield  {author} {\bibinfo {author} {\bibfnamefont {T.}~\bibnamefont
  {Zanca}}, \bibinfo {author} {\bibfnamefont {F.}~\bibnamefont {Pellegrini}},
  \bibinfo {author} {\bibfnamefont {G.~E.}\ \bibnamefont {Santoro}}, \ and\
  \bibinfo {author} {\bibfnamefont {E.}~\bibnamefont {Tosatti}},\ }\bibfield
  {title} {\enquote {\bibinfo {title} {Frictional lubricity enhanced by quantum
  mechanics},}\ }\href {\doibase 10.1073/pnas.1801144115} {\bibfield  {journal}
  {\bibinfo  {journal} {Proceedings of the National Academy of Sciences}\
  }\textbf {\bibinfo {volume} {115}},\ \bibinfo {pages} {3547--3550} (\bibinfo
  {year} {2018})}\BibitemShut {NoStop}%
\bibitem [{\citenamefont {Garc\'{\i}a-Mata}\ \emph {et~al.}(2007)\citenamefont
  {Garc\'{\i}a-Mata}, \citenamefont {Zhirov},\ and\ \citenamefont
  {Shepelyansky}}]{Garcia-Mata}%
  \BibitemOpen
  \bibfield  {author} {\bibinfo {author} {\bibfnamefont {I.}~\bibnamefont
  {Garc\'{\i}a-Mata}}, \bibinfo {author} {\bibfnamefont {O.~V.}\ \bibnamefont
  {Zhirov}}, \ and\ \bibinfo {author} {\bibfnamefont {D.~L.}\ \bibnamefont
  {Shepelyansky}},\ }\bibfield  {title} {\enquote {\bibinfo {title}
  {Frenkel-kontorova model with cold trapped ions},}\ }\href {\doibase
  10.1140/epjd/e2006-00220-2} {\bibfield  {journal} {\bibinfo  {journal} {Eur.
  Phys. J. D}\ }\textbf {\bibinfo {volume} {41}},\ \bibinfo {pages} {325--330}
  (\bibinfo {year} {2007})}\BibitemShut {NoStop}%
\bibitem [{\citenamefont {Pruttivarasin}\ \emph {et~al.}(2011)\citenamefont
  {Pruttivarasin}, \citenamefont {Ramm}, \citenamefont {Talukdar},
  \citenamefont {Kreuter},\ and\ \citenamefont
  {Häffner}}]{Pruttivarasin:2011}%
  \BibitemOpen
  \bibfield  {author} {\bibinfo {author} {\bibfnamefont {T.}~\bibnamefont
  {Pruttivarasin}}, \bibinfo {author} {\bibfnamefont {M.}~\bibnamefont {Ramm}},
  \bibinfo {author} {\bibfnamefont {I.}~\bibnamefont {Talukdar}}, \bibinfo
  {author} {\bibfnamefont {A.}~\bibnamefont {Kreuter}}, \ and\ \bibinfo
  {author} {\bibfnamefont {H.}~\bibnamefont {Häffner}},\ }\bibfield  {title}
  {\enquote {\bibinfo {title} {Trapped ions in optical lattices for probing
  oscillator chain models},}\ }\href {\doibase 10.1088/1367-2630/13/7/075012}
  {\bibfield  {journal} {\bibinfo  {journal} {New Journal of Physics}\ }\textbf
  {\bibinfo {volume} {13}},\ \bibinfo {pages} {075012} (\bibinfo {year}
  {2011})}\BibitemShut {NoStop}%
\bibitem [{\citenamefont {Cetina}\ \emph {et~al.}(2013)\citenamefont {Cetina},
  \citenamefont {Bylinskii}, \citenamefont {Karpa}, \citenamefont {Gangloff},
  \citenamefont {Beck}, \citenamefont {Ge}, \citenamefont {Scholz},
  \citenamefont {Grier}, \citenamefont {Chuang},\ and\ \citenamefont
  {Vuleti\'{c}}}]{Cetina:2013}%
  \BibitemOpen
  \bibfield  {author} {\bibinfo {author} {\bibfnamefont {M.}~\bibnamefont
  {Cetina}}, \bibinfo {author} {\bibfnamefont {A.}~\bibnamefont {Bylinskii}},
  \bibinfo {author} {\bibfnamefont {L.}~\bibnamefont {Karpa}}, \bibinfo
  {author} {\bibfnamefont {D.}~\bibnamefont {Gangloff}}, \bibinfo {author}
  {\bibfnamefont {K.~M.}\ \bibnamefont {Beck}}, \bibinfo {author}
  {\bibfnamefont {Y.}~\bibnamefont {Ge}}, \bibinfo {author} {\bibfnamefont
  {M.}~\bibnamefont {Scholz}}, \bibinfo {author} {\bibfnamefont {A.~T.}\
  \bibnamefont {Grier}}, \bibinfo {author} {\bibfnamefont {I.}~\bibnamefont
  {Chuang}}, \ and\ \bibinfo {author} {\bibfnamefont {V.}~\bibnamefont
  {Vuleti\'{c}}},\ }\bibfield  {title} {\enquote {\bibinfo {title}
  {One-dimensional array of ion chains coupled to an optical cavity},}\ }\href
  {\doibase 10.1088/1367-2630/15/5/053001} {\bibfield  {journal} {\bibinfo
  {journal} {New Journal of Physics}\ }\textbf {\bibinfo {volume} {15}},\
  \bibinfo {pages} {053001} (\bibinfo {year} {2013})}\BibitemShut {NoStop}%
\bibitem [{\citenamefont {Cormick}\ and\ \citenamefont
  {Morigi}(2013)}]{Cormick:2013}%
  \BibitemOpen
  \bibfield  {author} {\bibinfo {author} {\bibfnamefont {C.}~\bibnamefont
  {Cormick}}\ and\ \bibinfo {author} {\bibfnamefont {G.}~\bibnamefont
  {Morigi}},\ }\bibfield  {title} {\enquote {\bibinfo {title} {Ion chains in
  high-finesse cavities},}\ }\href {\doibase 10.1103/PhysRevA.87.013829}
  {\bibfield  {journal} {\bibinfo  {journal} {Phys. Rev. A}\ }\textbf {\bibinfo
  {volume} {87}},\ \bibinfo {pages} {013829} (\bibinfo {year}
  {2013})}\BibitemShut {NoStop}%
\bibitem [{\citenamefont {Braun}\ and\ \citenamefont
  {Kivshar}(2004)}]{Braun_Kishvar}%
  \BibitemOpen
  \bibfield  {author} {\bibinfo {author} {\bibfnamefont {O.M.}\ \bibnamefont
  {Braun}}\ and\ \bibinfo {author} {\bibfnamefont {Y.S.}\ \bibnamefont
  {Kivshar}},\ }\href@noop {} {\emph {\bibinfo {title} {The Frenkel-Kontorova
  Model: Concepts, Methods, and Applications}}}\ (\bibinfo  {publisher}
  {Springer},\ \bibinfo {address} {New York},\ \bibinfo {year}
  {2004})\BibitemShut {NoStop}%
\bibitem [{\citenamefont {Aubry}\ and\ \citenamefont {{Le
  Daeron}}(1983)}]{Aubry:1983}%
  \BibitemOpen
  \bibfield  {author} {\bibinfo {author} {\bibfnamefont {S.}~\bibnamefont
  {Aubry}}\ and\ \bibinfo {author} {\bibfnamefont {P.Y.}\ \bibnamefont {{Le
  Daeron}}},\ }\bibfield  {title} {\enquote {\bibinfo {title} {{The discrete
  Frenkel-Kontorova model and its extensions: I. Exact results for the
  ground-states}},}\ }\href {\doibase
  https://doi.org/10.1016/0167-2789(83)90233-6} {\bibfield  {journal} {\bibinfo
   {journal} {Physica D: Nonlinear Phenomena}\ }\textbf {\bibinfo {volume}
  {8}},\ \bibinfo {pages} {381--422} (\bibinfo {year} {1983})}\BibitemShut
  {NoStop}%
\bibitem [{\citenamefont {Gangloff}\ \emph {et~al.}(2015)\citenamefont
  {Gangloff}, \citenamefont {Bylinskii}, \citenamefont {Counts}, \citenamefont
  {Jhe},\ and\ \citenamefont {Vuleti\'c}}]{gangloff_velocity_2015}%
  \BibitemOpen
  \bibfield  {author} {\bibinfo {author} {\bibfnamefont {D.}~\bibnamefont
  {Gangloff}}, \bibinfo {author} {\bibfnamefont {A.}~\bibnamefont {Bylinskii}},
  \bibinfo {author} {\bibfnamefont {I.}~\bibnamefont {Counts}}, \bibinfo
  {author} {\bibfnamefont {W.}~\bibnamefont {Jhe}}, \ and\ \bibinfo {author}
  {\bibfnamefont {V.}~\bibnamefont {Vuleti\'c}},\ }\bibfield  {title} {\enquote
  {\bibinfo {title} {Velocity tuning of friction with two trapped atoms},}\
  }\href {\doibase 10.1038/nphys3459} {\bibfield  {journal} {\bibinfo
  {journal} {Nature Phys}\ }\textbf {\bibinfo {volume} {11}},\ \bibinfo {pages}
  {915--919} (\bibinfo {year} {2015})}\BibitemShut {NoStop}%
\bibitem [{\citenamefont {Bylinskii}\ \emph {et~al.}(2015)\citenamefont
  {Bylinskii}, \citenamefont {Gangloff},\ and\ \citenamefont
  {Vuleti{\'c}}}]{Bylinskii:2015}%
  \BibitemOpen
  \bibfield  {author} {\bibinfo {author} {\bibfnamefont {A.}~\bibnamefont
  {Bylinskii}}, \bibinfo {author} {\bibfnamefont {D.}~\bibnamefont {Gangloff}},
  \ and\ \bibinfo {author} {\bibfnamefont {V.}~\bibnamefont {Vuleti{\'c}}},\
  }\bibfield  {title} {\enquote {\bibinfo {title} {Tuning friction atom-by-atom
  in an ion-crystal simulator},}\ }\href {\doibase 10.1126/science.1261422}
  {\bibfield  {journal} {\bibinfo  {journal} {Science}\ }\textbf {\bibinfo
  {volume} {348}},\ \bibinfo {pages} {1115--1118} (\bibinfo {year}
  {2015})}\BibitemShut {NoStop}%
\bibitem [{\citenamefont {Bylinskii}\ \emph {et~al.}(2016)\citenamefont
  {Bylinskii}, \citenamefont {Gangloff}, \citenamefont {Counts},\ and\
  \citenamefont {Vuleti{\'c}}}]{Bylinskii:2016}%
  \BibitemOpen
  \bibfield  {author} {\bibinfo {author} {\bibfnamefont {A.}~\bibnamefont
  {Bylinskii}}, \bibinfo {author} {\bibfnamefont {D.}~\bibnamefont {Gangloff}},
  \bibinfo {author} {\bibfnamefont {I.}~\bibnamefont {Counts}}, \ and\ \bibinfo
  {author} {\bibfnamefont {V.}~\bibnamefont {Vuleti{\'c}}},\ }\bibfield
  {title} {\enquote {\bibinfo {title} {{Observation of Aubry-type transition in
  finite atom chains via friction}},}\ }\href {\doibase 10.1038/nmat4601}
  {\bibfield  {journal} {\bibinfo  {journal} {Nature Materials}\ }\textbf
  {\bibinfo {volume} {15}},\ \bibinfo {pages} {717--721} (\bibinfo {year}
  {2016})}\BibitemShut {NoStop}%
\bibitem [{\citenamefont {Linnet}\ \emph {et~al.}(2012)\citenamefont {Linnet},
  \citenamefont {Leroux}, \citenamefont {Marciante}, \citenamefont {Dantan},\
  and\ \citenamefont {Drewsen}}]{Linnet:2012}%
  \BibitemOpen
  \bibfield  {author} {\bibinfo {author} {\bibfnamefont {R.~B.}\ \bibnamefont
  {Linnet}}, \bibinfo {author} {\bibfnamefont {I.~D.}\ \bibnamefont {Leroux}},
  \bibinfo {author} {\bibfnamefont {M.}~\bibnamefont {Marciante}}, \bibinfo
  {author} {\bibfnamefont {A.}~\bibnamefont {Dantan}}, \ and\ \bibinfo {author}
  {\bibfnamefont {M.}~\bibnamefont {Drewsen}},\ }\bibfield  {title} {\enquote
  {\bibinfo {title} {Pinning an ion with an intracavity optical lattice},}\
  }\href {\doibase 10.1103/PhysRevLett.109.233005} {\bibfield  {journal}
  {\bibinfo  {journal} {Phys. Rev. Lett.}\ }\textbf {\bibinfo {volume} {109}},\
  \bibinfo {pages} {233005} (\bibinfo {year} {2012})}\BibitemShut {NoStop}%
\bibitem [{\citenamefont {Lechner}\ \emph {et~al.}(2016)\citenamefont
  {Lechner}, \citenamefont {Maier}, \citenamefont {Hempel}, \citenamefont
  {Jurcevic}, \citenamefont {Lanyon}, \citenamefont {Monz}, \citenamefont
  {Brownnutt}, \citenamefont {Blatt},\ and\ \citenamefont
  {Roos}}]{Lechner:2016}%
  \BibitemOpen
  \bibfield  {author} {\bibinfo {author} {\bibfnamefont {R.}~\bibnamefont
  {Lechner}}, \bibinfo {author} {\bibfnamefont {C.}~\bibnamefont {Maier}},
  \bibinfo {author} {\bibfnamefont {C.}~\bibnamefont {Hempel}}, \bibinfo
  {author} {\bibfnamefont {P.}~\bibnamefont {Jurcevic}}, \bibinfo {author}
  {\bibfnamefont {B.~P.}\ \bibnamefont {Lanyon}}, \bibinfo {author}
  {\bibfnamefont {T.}~\bibnamefont {Monz}}, \bibinfo {author} {\bibfnamefont
  {M.}~\bibnamefont {Brownnutt}}, \bibinfo {author} {\bibfnamefont
  {R.}~\bibnamefont {Blatt}}, \ and\ \bibinfo {author} {\bibfnamefont {C.~F.}\
  \bibnamefont {Roos}},\ }\bibfield  {title} {\enquote {\bibinfo {title}
  {Electromagnetically-induced-transparency ground-state cooling of long ion
  strings},}\ }\href {\doibase 10.1103/PhysRevA.93.053401} {\bibfield
  {journal} {\bibinfo  {journal} {Phys. Rev. A}\ }\textbf {\bibinfo {volume}
  {93}},\ \bibinfo {pages} {053401} (\bibinfo {year} {2016})}\BibitemShut
  {NoStop}%
\bibitem [{\citenamefont {Feng}\ \emph {et~al.}(2020)\citenamefont {Feng},
  \citenamefont {Tan}, \citenamefont {De}, \citenamefont {Menon}, \citenamefont
  {Chu}, \citenamefont {Pagano},\ and\ \citenamefont {Monroe}}]{Feng:2020}%
  \BibitemOpen
  \bibfield  {author} {\bibinfo {author} {\bibfnamefont {L.}~\bibnamefont
  {Feng}}, \bibinfo {author} {\bibfnamefont {W.~L.}\ \bibnamefont {Tan}},
  \bibinfo {author} {\bibfnamefont {A.}~\bibnamefont {De}}, \bibinfo {author}
  {\bibfnamefont {A.}~\bibnamefont {Menon}}, \bibinfo {author} {\bibfnamefont
  {A.}~\bibnamefont {Chu}}, \bibinfo {author} {\bibfnamefont {G.}~\bibnamefont
  {Pagano}}, \ and\ \bibinfo {author} {\bibfnamefont {C.}~\bibnamefont
  {Monroe}},\ }\bibfield  {title} {\enquote {\bibinfo {title} {{Efficient
  Ground-State Cooling of Large Trapped-Ion Chains with an
  Electromagnetically-Induced-Transparency Tripod Scheme}},}\ }\href {\doibase
  10.1103/PhysRevLett.125.053001} {\bibfield  {journal} {\bibinfo  {journal}
  {Phys. Rev. Lett.}\ }\textbf {\bibinfo {volume} {125}},\ \bibinfo {pages}
  {053001} (\bibinfo {year} {2020})}\BibitemShut {NoStop}%
\bibitem [{\citenamefont {Schmidt}\ \emph {et~al.}(2018)\citenamefont
  {Schmidt}, \citenamefont {Lambrecht}, \citenamefont {Weckesser},
  \citenamefont {Debatin}, \citenamefont {Karpa},\ and\ \citenamefont
  {Schaetz}}]{Schmidt:2018}%
  \BibitemOpen
  \bibfield  {author} {\bibinfo {author} {\bibfnamefont {J.}~\bibnamefont
  {Schmidt}}, \bibinfo {author} {\bibfnamefont {A.}~\bibnamefont {Lambrecht}},
  \bibinfo {author} {\bibfnamefont {P.}~\bibnamefont {Weckesser}}, \bibinfo
  {author} {\bibfnamefont {M.}~\bibnamefont {Debatin}}, \bibinfo {author}
  {\bibfnamefont {L.}~\bibnamefont {Karpa}}, \ and\ \bibinfo {author}
  {\bibfnamefont {T.}~\bibnamefont {Schaetz}},\ }\bibfield  {title} {\enquote
  {\bibinfo {title} {{Optical Trapping of Ion Coulomb Crystals}},}\ }\href
  {\doibase 10.1103/PhysRevX.8.021028} {\bibfield  {journal} {\bibinfo
  {journal} {Phys. Rev. X}\ }\textbf {\bibinfo {volume} {8}},\ \bibinfo {pages}
  {021028} (\bibinfo {year} {2018})}\BibitemShut {NoStop}%
\bibitem [{\citenamefont {Hoenig}\ \emph {et~al.}(2023)\citenamefont {Hoenig},
  \citenamefont {Thielemann}, \citenamefont {Karpa}, \citenamefont {Walker},
  \citenamefont {Mohammadi},\ and\ \citenamefont {Schaetz}}]{Hoenig:2023}%
  \BibitemOpen
  \bibfield  {author} {\bibinfo {author} {\bibfnamefont {D.}~\bibnamefont
  {Hoenig}}, \bibinfo {author} {\bibfnamefont {F.}~\bibnamefont {Thielemann}},
  \bibinfo {author} {\bibfnamefont {L.}~\bibnamefont {Karpa}}, \bibinfo
  {author} {\bibfnamefont {T.}~\bibnamefont {Walker}}, \bibinfo {author}
  {\bibfnamefont {A.}~\bibnamefont {Mohammadi}}, \ and\ \bibinfo {author}
  {\bibfnamefont {T.}~\bibnamefont {Schaetz}},\ }\bibfield  {title} {\enquote
  {\bibinfo {title} {{Trapping Ion Coulomb Crystals in an Optical Lattice}},}\
  }\href@noop {} {\bibfield  {journal} {\bibinfo  {journal} {arXiv:2306.12518}\
  } (\bibinfo {year} {2023})}\BibitemShut {NoStop}%
\bibitem [{\citenamefont {Pokrovsky}\ and\ \citenamefont
  {Virosztek}(1983)}]{Pokrovsky_1983}%
  \BibitemOpen
  \bibfield  {author} {\bibinfo {author} {\bibfnamefont {V~L}\ \bibnamefont
  {Pokrovsky}}\ and\ \bibinfo {author} {\bibfnamefont {A}~\bibnamefont
  {Virosztek}},\ }\bibfield  {title} {\enquote {\bibinfo {title} {Long-range
  interactions in commensurate-incommensurate phase transition},}\ }\href
  {\doibase 10.1088/0022-3719/16/23/013} {\bibfield  {journal} {\bibinfo
  {journal} {Journal of Physics C: Solid State Physics}\ }\textbf {\bibinfo
  {volume} {16}},\ \bibinfo {pages} {4513--4525} (\bibinfo {year}
  {1983})}\BibitemShut {NoStop}%
\bibitem [{\citenamefont {Frank}\ \emph {et~al.}(1949)\citenamefont {Frank},
  \citenamefont {van~der Merwe},\ and\ \citenamefont {Mott}}]{Merwe}%
  \BibitemOpen
  \bibfield  {author} {\bibinfo {author} {\bibfnamefont {F.~C.}\ \bibnamefont
  {Frank}}, \bibinfo {author} {\bibfnamefont {J.~H.}\ \bibnamefont {van~der
  Merwe}}, \ and\ \bibinfo {author} {\bibfnamefont {Nevill~Francis}\
  \bibnamefont {Mott}},\ }\bibfield  {title} {\enquote {\bibinfo {title}
  {One-dimensional dislocations. i. static theory},}\ }\href {\doibase
  10.1098/rspa.1949.0095} {\bibfield  {journal} {\bibinfo  {journal}
  {Proceedings of the Royal Society of London. Series A. Mathematical and
  Physical Sciences}\ }\textbf {\bibinfo {volume} {198}},\ \bibinfo {pages}
  {205--216} (\bibinfo {year} {1949})}\BibitemShut {NoStop}%
\bibitem [{\citenamefont {Landa}\ \emph {et~al.}(2020)\citenamefont {Landa},
  \citenamefont {Cormick},\ and\ \citenamefont {Morigi}}]{Landa:2020}%
  \BibitemOpen
  \bibfield  {author} {\bibinfo {author} {\bibfnamefont {H.}~\bibnamefont
  {Landa}}, \bibinfo {author} {\bibfnamefont {C.}~\bibnamefont {Cormick}}, \
  and\ \bibinfo {author} {\bibfnamefont {G.}~\bibnamefont {Morigi}},\
  }\bibfield  {title} {\enquote {\bibinfo {title} {Static kinks in chains of
  interacting atoms},}\ }\href {\doibase 10.3390/condmat5020035} {\bibfield
  {journal} {\bibinfo  {journal} {Condensed Matter}\ }\textbf {\bibinfo
  {volume} {5}} (\bibinfo {year} {2020}),\ 10.3390/condmat5020035}\BibitemShut
  {NoStop}%
\bibitem [{\citenamefont {Menu}\ \emph {et~al.}(2024)\citenamefont {Menu},
  \citenamefont {Yago~Malo}, \citenamefont {Vuleti\ifmmode~\acute{c}\else
  \'{c}\fi{}}, \citenamefont {Chiofalo},\ and\ \citenamefont
  {Morigi}}]{Menu:2023}%
  \BibitemOpen
  \bibfield  {author} {\bibinfo {author} {\bibfnamefont {R.}~\bibnamefont
  {Menu}}, \bibinfo {author} {\bibfnamefont {J.}~\bibnamefont {Yago~Malo}},
  \bibinfo {author} {\bibfnamefont {V.}~\bibnamefont
  {Vuleti\ifmmode~\acute{c}\else \'{c}\fi{}}}, \bibinfo {author} {\bibfnamefont
  {M.~L.}\ \bibnamefont {Chiofalo}}, \ and\ \bibinfo {author} {\bibfnamefont
  {G.}~\bibnamefont {Morigi}},\ }\bibfield  {title} {\enquote {\bibinfo {title}
  {{Quantum frustrated Wigner chains}},}\ }\href {\doibase
  10.1103/PhysRevB.110.155121} {\bibfield  {journal} {\bibinfo  {journal}
  {Phys. Rev. B}\ }\textbf {\bibinfo {volume} {110}},\ \bibinfo {pages}
  {155121} (\bibinfo {year} {2024})}\BibitemShut {NoStop}%
\bibitem [{\citenamefont {Landa}\ \emph {et~al.}(2013)\citenamefont {Landa},
  \citenamefont {Reznik}, \citenamefont {Brox}, \citenamefont {Mielenz},\ and\
  \citenamefont {Schaetz}}]{Landa:2013}%
  \BibitemOpen
  \bibfield  {author} {\bibinfo {author} {\bibfnamefont {H.}~\bibnamefont
  {Landa}}, \bibinfo {author} {\bibfnamefont {B.}~\bibnamefont {Reznik}},
  \bibinfo {author} {\bibfnamefont {J.}~\bibnamefont {Brox}}, \bibinfo {author}
  {\bibfnamefont {M.}~\bibnamefont {Mielenz}}, \ and\ \bibinfo {author}
  {\bibfnamefont {T.}~\bibnamefont {Schaetz}},\ }\bibfield  {title} {\enquote
  {\bibinfo {title} {Structure, dynamics and bifurcations of discrete solitons
  in trapped ion crystals},}\ }\href {\doibase 10.1088/1367-2630/15/9/093003}
  {\bibfield  {journal} {\bibinfo  {journal} {New Journal of Physics}\ }\textbf
  {\bibinfo {volume} {15}},\ \bibinfo {pages} {093003} (\bibinfo {year}
  {2013})}\BibitemShut {NoStop}%
\bibitem [{\citenamefont {Partner}\ \emph {et~al.}(2013)\citenamefont
  {Partner}, \citenamefont {Nigmatullin}, \citenamefont {Burgermeister},
  \citenamefont {Pyka}, \citenamefont {Keller}, \citenamefont {Retzker},
  \citenamefont {Plenio},\ and\ \citenamefont {Mehlst\"aubler}}]{Partner:2013}%
  \BibitemOpen
  \bibfield  {author} {\bibinfo {author} {\bibfnamefont {H.~L.}\ \bibnamefont
  {Partner}}, \bibinfo {author} {\bibfnamefont {R.}~\bibnamefont
  {Nigmatullin}}, \bibinfo {author} {\bibfnamefont {T.}~\bibnamefont
  {Burgermeister}}, \bibinfo {author} {\bibfnamefont {K.}~\bibnamefont {Pyka}},
  \bibinfo {author} {\bibfnamefont {J.}~\bibnamefont {Keller}}, \bibinfo
  {author} {\bibfnamefont {A.}~\bibnamefont {Retzker}}, \bibinfo {author}
  {\bibfnamefont {M.~B.}\ \bibnamefont {Plenio}}, \ and\ \bibinfo {author}
  {\bibfnamefont {T.~E.}\ \bibnamefont {Mehlst\"aubler}},\ }\bibfield  {title}
  {\enquote {\bibinfo {title} {{Dynamics of topological defects in ion Coulomb
  crystals}},}\ }\href {\doibase 10.1088/1367-2630/15/10/103013} {\bibfield
  {journal} {\bibinfo  {journal} {New Journal of Physics}\ }\textbf {\bibinfo
  {volume} {15}},\ \bibinfo {pages} {103013} (\bibinfo {year}
  {2013})}\BibitemShut {NoStop}%
\bibitem [{\citenamefont {Willis}\ \emph {et~al.}(1986)\citenamefont {Willis},
  \citenamefont {El-Batanouny},\ and\ \citenamefont {Stancioff}}]{Willis:1986}%
  \BibitemOpen
  \bibfield  {author} {\bibinfo {author} {\bibfnamefont {C.}~\bibnamefont
  {Willis}}, \bibinfo {author} {\bibfnamefont {M.}~\bibnamefont
  {El-Batanouny}}, \ and\ \bibinfo {author} {\bibfnamefont {P.}~\bibnamefont
  {Stancioff}},\ }\bibfield  {title} {\enquote {\bibinfo {title} {{Sine-Gordon
  kinks on a discrete lattice. I. Hamiltonian formalism}},}\ }\href {\doibase
  10.1103/PhysRevB.33.1904} {\bibfield  {journal} {\bibinfo  {journal} {Phys.
  Rev. B}\ }\textbf {\bibinfo {volume} {33}},\ \bibinfo {pages} {1904--1911}
  (\bibinfo {year} {1986})}\BibitemShut {NoStop}%
\bibitem [{\citenamefont {Braun}\ \emph {et~al.}(1990)\citenamefont {Braun},
  \citenamefont {Kivshar},\ and\ \citenamefont {Zelenskaya}}]{Braun:1990}%
  \BibitemOpen
  \bibfield  {author} {\bibinfo {author} {\bibfnamefont {O.~M.}\ \bibnamefont
  {Braun}}, \bibinfo {author} {\bibfnamefont {Yu.~S.}\ \bibnamefont {Kivshar}},
  \ and\ \bibinfo {author} {\bibfnamefont {I.~I.}\ \bibnamefont {Zelenskaya}},\
  }\bibfield  {title} {\enquote {\bibinfo {title} {{Kinks in the
  Frenkel-Kontorova model with long-range interparticle interactions}},}\
  }\href {\doibase 10.1103/PhysRevB.41.7118} {\bibfield  {journal} {\bibinfo
  {journal} {Phys. Rev. B}\ }\textbf {\bibinfo {volume} {41}},\ \bibinfo
  {pages} {7118--7138} (\bibinfo {year} {1990})}\BibitemShut {NoStop}%
\bibitem [{\citenamefont {Chelpanova}\ \emph {et~al.}(2024)\citenamefont
  {Chelpanova}, \citenamefont {Kelly}, \citenamefont {Schmidt-Kaler},
  \citenamefont {Morigi},\ and\ \citenamefont {Marino}}]{Chelpanova:2024}%
  \BibitemOpen
  \bibfield  {author} {\bibinfo {author} {\bibfnamefont {O.}~\bibnamefont
  {Chelpanova}}, \bibinfo {author} {\bibfnamefont {S.~P.}\ \bibnamefont
  {Kelly}}, \bibinfo {author} {\bibfnamefont {F.}~\bibnamefont
  {Schmidt-Kaler}}, \bibinfo {author} {\bibfnamefont {G.}~\bibnamefont
  {Morigi}}, \ and\ \bibinfo {author} {\bibfnamefont {J.}~\bibnamefont
  {Marino}},\ }\href@noop {} {\enquote {\bibinfo {title} {Dynamics of quantum
  discommensurations in the frenkel-kontorova chain},}\ } (\bibinfo {year}
  {2024}),\ \Eprint {http://arxiv.org/abs/2401.12614} {arXiv:2401.12614
  [cond-mat.stat-mech]} \BibitemShut {NoStop}%
\bibitem [{\citenamefont {Hubbard}(1978)}]{Hubbard}%
  \BibitemOpen
  \bibfield  {author} {\bibinfo {author} {\bibfnamefont {J.}~\bibnamefont
  {Hubbard}},\ }\bibfield  {title} {\enquote {\bibinfo {title} {Generalized
  wigner lattices in one dimension and some applications to
  tetracyanoquinodimethane (tcnq) salts},}\ }\href {\doibase
  10.1103/PhysRevB.17.494} {\bibfield  {journal} {\bibinfo  {journal} {Phys.
  Rev. B}\ }\textbf {\bibinfo {volume} {17}},\ \bibinfo {pages} {494--505}
  (\bibinfo {year} {1978})}\BibitemShut {NoStop}%
\bibitem [{\citenamefont {Beyeler}\ \emph {et~al.}(1980)\citenamefont
  {Beyeler}, \citenamefont {Pietronero},\ and\ \citenamefont
  {Str\"assler}}]{Pietronero}%
  \BibitemOpen
  \bibfield  {author} {\bibinfo {author} {\bibfnamefont {H.~U.}\ \bibnamefont
  {Beyeler}}, \bibinfo {author} {\bibfnamefont {L.}~\bibnamefont {Pietronero}},
  \ and\ \bibinfo {author} {\bibfnamefont {S.}~\bibnamefont {Str\"assler}},\
  }\bibfield  {title} {\enquote {\bibinfo {title} {Configurational model for a
  one-dimensional ionic conductor},}\ }\href {\doibase
  10.1103/PhysRevB.22.2988} {\bibfield  {journal} {\bibinfo  {journal} {Phys.
  Rev. B}\ }\textbf {\bibinfo {volume} {22}},\ \bibinfo {pages} {2988--3000}
  (\bibinfo {year} {1980})}\BibitemShut {NoStop}%
\bibitem [{\citenamefont {Dalmonte}\ \emph {et~al.}(2010)\citenamefont
  {Dalmonte}, \citenamefont {Pupillo},\ and\ \citenamefont
  {Zoller}}]{Dalmonte:2010}%
  \BibitemOpen
  \bibfield  {author} {\bibinfo {author} {\bibfnamefont {M.}~\bibnamefont
  {Dalmonte}}, \bibinfo {author} {\bibfnamefont {G.}~\bibnamefont {Pupillo}}, \
  and\ \bibinfo {author} {\bibfnamefont {P.}~\bibnamefont {Zoller}},\
  }\bibfield  {title} {\enquote {\bibinfo {title} {{One-Dimensional Quantum
  Liquids with Power-Law Interactions: The Luttinger Staircase}},}\ }\href
  {\doibase 10.1103/PhysRevLett.105.140401} {\bibfield  {journal} {\bibinfo
  {journal} {Phys. Rev. Lett.}\ }\textbf {\bibinfo {volume} {105}},\ \bibinfo
  {pages} {140401} (\bibinfo {year} {2010})}\BibitemShut {NoStop}%
\bibitem [{\citenamefont {Roux}\ \emph {et~al.}(2008)\citenamefont {Roux},
  \citenamefont {Barthel}, \citenamefont {McCulloch}, \citenamefont {Kollath},
  \citenamefont {Schollw\"ock},\ and\ \citenamefont {Giamarchi}}]{Roux:2008}%
  \BibitemOpen
  \bibfield  {author} {\bibinfo {author} {\bibfnamefont {G.}~\bibnamefont
  {Roux}}, \bibinfo {author} {\bibfnamefont {T.}~\bibnamefont {Barthel}},
  \bibinfo {author} {\bibfnamefont {I.~P.}\ \bibnamefont {McCulloch}}, \bibinfo
  {author} {\bibfnamefont {C.}~\bibnamefont {Kollath}}, \bibinfo {author}
  {\bibfnamefont {U.}~\bibnamefont {Schollw\"ock}}, \ and\ \bibinfo {author}
  {\bibfnamefont {T.}~\bibnamefont {Giamarchi}},\ }\bibfield  {title} {\enquote
  {\bibinfo {title} {Quasiperiodic bose-hubbard model and localization in
  one-dimensional cold atomic gases},}\ }\href {\doibase
  10.1103/PhysRevA.78.023628} {\bibfield  {journal} {\bibinfo  {journal} {Phys.
  Rev. A}\ }\textbf {\bibinfo {volume} {78}},\ \bibinfo {pages} {023628}
  (\bibinfo {year} {2008})}\BibitemShut {NoStop}%
\bibitem [{\citenamefont {Bak}\ and\ \citenamefont
  {Bruinsma}(1982)}]{Bak-PRL:1982}%
  \BibitemOpen
  \bibfield  {author} {\bibinfo {author} {\bibfnamefont {P.}~\bibnamefont
  {Bak}}\ and\ \bibinfo {author} {\bibfnamefont {R.}~\bibnamefont {Bruinsma}},\
  }\bibfield  {title} {\enquote {\bibinfo {title} {One-dimensional ising model
  and the complete devil's staircase},}\ }\href {\doibase
  10.1103/PhysRevLett.49.249} {\bibfield  {journal} {\bibinfo  {journal} {Phys.
  Rev. Lett.}\ }\textbf {\bibinfo {volume} {49}},\ \bibinfo {pages} {249--251}
  (\bibinfo {year} {1982})}\BibitemShut {NoStop}%
\bibitem [{\citenamefont {Aubry}(1983)}]{Aubry:1983a}%
  \BibitemOpen
  \bibfield  {author} {\bibinfo {author} {\bibfnamefont {S.}~\bibnamefont
  {Aubry}},\ }\bibfield  {title} {\enquote {\bibinfo {title} {{The twist map,
  the extended Frenkel-Kontorova model and the devil's staircase}},}\ }\href
  {\doibase https://doi.org/10.1016/0167-2789(83)90129-X} {\bibfield  {journal}
  {\bibinfo  {journal} {Physica D: Nonlinear Phenomena}\ }\textbf {\bibinfo
  {volume} {7}},\ \bibinfo {pages} {240--258} (\bibinfo {year}
  {1983})}\BibitemShut {NoStop}%
\bibitem [{\citenamefont {Krajewski}\ and\ \citenamefont
  {M{\"u}ser}(2005)}]{Mueser:2005}%
  \BibitemOpen
  \bibfield  {author} {\bibinfo {author} {\bibfnamefont {F.~R.}\ \bibnamefont
  {Krajewski}}\ and\ \bibinfo {author} {\bibfnamefont {M.~H.}\ \bibnamefont
  {M{\"u}ser}},\ }\bibfield  {title} {\enquote {\bibinfo {title} {{Quantum
  dynamics in the highly discrete, commensurate Frenkel Kontorova model: a
  path-integral molecular dynamics study.}}}\ }\href@noop {} {\bibfield
  {journal} {\bibinfo  {journal} {The Journal of chemical physics}\ }\textbf
  {\bibinfo {volume} {122 12}},\ \bibinfo {pages} {124711} (\bibinfo {year}
  {2005})}\BibitemShut {NoStop}%
\bibitem [{Note1()}]{Note1}%
  \BibitemOpen
  \bibinfo {note} {We note that, in the nearest-neighbour limit $\alpha \to
  \infty $, this expression reduces to Eq.\ (3.17) of Ref. \cite {Pietronero}.
  In order to perform a systematic comparison, we note that the coefficient $A$
  of Ref. \cite {Pietronero} corresponds to our coefficient $\protect \mathcal
  K$ and their coefficient $g$ is twice the coefficient $g$ of Eq. \protect
  \eqref {Eq:g}. With these substitutions Eq. \protect \eqref {Eq:Qq:1}
  coincides with Eq. (3.17) of Ref. \cite {Pietronero}.}\BibitemShut {Stop}%
\bibitem [{\citenamefont {Koziol}\ \emph {et~al.}(2023)\citenamefont {Koziol},
  \citenamefont {Duft}, \citenamefont {Morigi},\ and\ \citenamefont
  {Schmidt}}]{Koziol:2023}%
  \BibitemOpen
  \bibfield  {author} {\bibinfo {author} {\bibfnamefont {J.~A.}\ \bibnamefont
  {Koziol}}, \bibinfo {author} {\bibfnamefont {A.}~\bibnamefont {Duft}},
  \bibinfo {author} {\bibfnamefont {G.}~\bibnamefont {Morigi}}, \ and\ \bibinfo
  {author} {\bibfnamefont {K.~P.}\ \bibnamefont {Schmidt}},\ }\bibfield
  {title} {\enquote {\bibinfo {title} {{Systematic analysis of crystalline
  phases in bosonic lattice models with algebraically decaying density-density
  interactions}},}\ }\href {\doibase 10.21468/SciPostPhys.14.5.136} {\bibfield
  {journal} {\bibinfo  {journal} {SciPost Phys.}\ }\textbf {\bibinfo {volume}
  {14}},\ \bibinfo {pages} {136} (\bibinfo {year} {2023})}\BibitemShut
  {NoStop}%
\bibitem [{\citenamefont {Bak}(1982)}]{Bak_1982}%
  \BibitemOpen
  \bibfield  {author} {\bibinfo {author} {\bibfnamefont {P.}~\bibnamefont
  {Bak}},\ }\bibfield  {title} {\enquote {\bibinfo {title} {Commensurate
  phases, incommensurate phases and the devil's staircase},}\ }\href {\doibase
  10.1088/0034-4885/45/6/001} {\bibfield  {journal} {\bibinfo  {journal}
  {Reports on Progress in Physics}\ }\textbf {\bibinfo {volume} {45}},\
  \bibinfo {pages} {587--629} (\bibinfo {year} {1982})}\BibitemShut {NoStop}%
\bibitem [{\citenamefont {Campa}\ \emph {et~al.}(2009)\citenamefont {Campa},
  \citenamefont {Dauxois},\ and\ \citenamefont {Ruffo}}]{CAMPA200957}%
  \BibitemOpen
  \bibfield  {author} {\bibinfo {author} {\bibfnamefont {A.}~\bibnamefont
  {Campa}}, \bibinfo {author} {\bibfnamefont {T.}~\bibnamefont {Dauxois}}, \
  and\ \bibinfo {author} {\bibfnamefont {S.}~\bibnamefont {Ruffo}},\ }\bibfield
   {title} {\enquote {\bibinfo {title} {Statistical mechanics and dynamics of
  solvable models with long-range interactions},}\ }\href {\doibase
  https://doi.org/10.1016/j.physrep.2009.07.001} {\bibfield  {journal}
  {\bibinfo  {journal} {Physics Reports}\ }\textbf {\bibinfo {volume} {480}},\
  \bibinfo {pages} {57--159} (\bibinfo {year} {2009})}\BibitemShut {NoStop}%
\bibitem [{\citenamefont {Defenu}\ \emph {et~al.}(2019)\citenamefont {Defenu},
  \citenamefont {Morigi}, \citenamefont {Dell'Anna},\ and\ \citenamefont
  {Enss}}]{Defenu:2019}%
  \BibitemOpen
  \bibfield  {author} {\bibinfo {author} {\bibfnamefont {N.}~\bibnamefont
  {Defenu}}, \bibinfo {author} {\bibfnamefont {G.}~\bibnamefont {Morigi}},
  \bibinfo {author} {\bibfnamefont {L.}~\bibnamefont {Dell'Anna}}, \ and\
  \bibinfo {author} {\bibfnamefont {T.}~\bibnamefont {Enss}},\ }\bibfield
  {title} {\enquote {\bibinfo {title} {{Universal dynamical scaling of
  long-range topological superconductors}},}\ }\href {\doibase
  10.1103/PhysRevB.100.184306} {\bibfield  {journal} {\bibinfo  {journal}
  {Phys. Rev. B}\ }\textbf {\bibinfo {volume} {100}},\ \bibinfo {pages}
  {184306} (\bibinfo {year} {2019})}\BibitemShut {NoStop}%
\bibitem [{\citenamefont {Bruinsma}\ and\ \citenamefont
  {Bak}(1983)}]{Bruinsma:1983}%
  \BibitemOpen
  \bibfield  {author} {\bibinfo {author} {\bibfnamefont {R.}~\bibnamefont
  {Bruinsma}}\ and\ \bibinfo {author} {\bibfnamefont {P.}~\bibnamefont {Bak}},\
  }\bibfield  {title} {\enquote {\bibinfo {title} {Self-similarity and fractal
  dimension of the devil's staircase in the one-dimensional ising model},}\
  }\href {\doibase 10.1103/PhysRevB.27.5824} {\bibfield  {journal} {\bibinfo
  {journal} {Phys. Rev. B}\ }\textbf {\bibinfo {volume} {27}},\ \bibinfo
  {pages} {5824--5825} (\bibinfo {year} {1983})}\BibitemShut {NoStop}%
\bibitem [{\citenamefont {Peierls}(1936)}]{Peierls}%
  \BibitemOpen
  \bibfield  {author} {\bibinfo {author} {\bibfnamefont {R.}~\bibnamefont
  {Peierls}},\ }\bibfield  {title} {\enquote {\bibinfo {title} {{On Ising’s
  model of ferromagnetism}},}\ }\href {\doibase 10.1017/S0305004100019174}
  {\bibfield  {journal} {\bibinfo  {journal} {Mathematical Proceedings of the
  Cambridge Philosophical Society}\ }\textbf {\bibinfo {volume} {32}},\
  \bibinfo {pages} {477–481} (\bibinfo {year} {1936})}\BibitemShut {NoStop}%
\bibitem [{\citenamefont {Erba}\ \emph {et~al.}(2021)\citenamefont {Erba},
  \citenamefont {Pastore},\ and\ \citenamefont
  {Rotondo}}]{PhysRevLett.126.183601}%
  \BibitemOpen
  \bibfield  {author} {\bibinfo {author} {\bibfnamefont {V.}~\bibnamefont
  {Erba}}, \bibinfo {author} {\bibfnamefont {M.}~\bibnamefont {Pastore}}, \
  and\ \bibinfo {author} {\bibfnamefont {P.}~\bibnamefont {Rotondo}},\
  }\bibfield  {title} {\enquote {\bibinfo {title} {{Self-Induced Glassy Phase
  in Multimodal Cavity Quantum Electrodynamics}},}\ }\href {\doibase
  10.1103/PhysRevLett.126.183601} {\bibfield  {journal} {\bibinfo  {journal}
  {Phys. Rev. Lett.}\ }\textbf {\bibinfo {volume} {126}},\ \bibinfo {pages}
  {183601} (\bibinfo {year} {2021})}\BibitemShut {NoStop}%
\bibitem [{\citenamefont {Li}\ \emph {et~al.}(2017)\citenamefont {Li},
  \citenamefont {Urban}, \citenamefont {Noel}, \citenamefont {Chuang},
  \citenamefont {Xia}, \citenamefont {Ransford}, \citenamefont {Hemmerling},
  \citenamefont {Wang}, \citenamefont {Li}, \citenamefont {H\"affner},\ and\
  \citenamefont {Zhang}}]{Haeffner}%
  \BibitemOpen
  \bibfield  {author} {\bibinfo {author} {\bibfnamefont {H.-K.}\ \bibnamefont
  {Li}}, \bibinfo {author} {\bibfnamefont {E.}~\bibnamefont {Urban}}, \bibinfo
  {author} {\bibfnamefont {C.}~\bibnamefont {Noel}}, \bibinfo {author}
  {\bibfnamefont {A.}~\bibnamefont {Chuang}}, \bibinfo {author} {\bibfnamefont
  {Y.}~\bibnamefont {Xia}}, \bibinfo {author} {\bibfnamefont {A.}~\bibnamefont
  {Ransford}}, \bibinfo {author} {\bibfnamefont {B.}~\bibnamefont
  {Hemmerling}}, \bibinfo {author} {\bibfnamefont {Y.}~\bibnamefont {Wang}},
  \bibinfo {author} {\bibfnamefont {T.}~\bibnamefont {Li}}, \bibinfo {author}
  {\bibfnamefont {H.}~\bibnamefont {H\"affner}}, \ and\ \bibinfo {author}
  {\bibfnamefont {X.}~\bibnamefont {Zhang}},\ }\bibfield  {title} {\enquote
  {\bibinfo {title} {{Realization of Translational Symmetry in Trapped Cold Ion
  Rings}},}\ }\href {\doibase 10.1103/PhysRevLett.118.053001} {\bibfield
  {journal} {\bibinfo  {journal} {Phys. Rev. Lett.}\ }\textbf {\bibinfo
  {volume} {118}},\ \bibinfo {pages} {053001} (\bibinfo {year}
  {2017})}\BibitemShut {NoStop}%
\bibitem [{\citenamefont {Landa}\ \emph {et~al.}(2014)\citenamefont {Landa},
  \citenamefont {Retzker}, \citenamefont {Schaetz},\ and\ \citenamefont
  {Reznik}}]{Landa:2014}%
  \BibitemOpen
  \bibfield  {author} {\bibinfo {author} {\bibfnamefont {H.}~\bibnamefont
  {Landa}}, \bibinfo {author} {\bibfnamefont {A.}~\bibnamefont {Retzker}},
  \bibinfo {author} {\bibfnamefont {T.}~\bibnamefont {Schaetz}}, \ and\
  \bibinfo {author} {\bibfnamefont {B.}~\bibnamefont {Reznik}},\ }\bibfield
  {title} {\enquote {\bibinfo {title} {Entanglement generation using discrete
  solitons in coulomb crystals},}\ }\href {\doibase
  10.1103/PhysRevLett.113.053001} {\bibfield  {journal} {\bibinfo  {journal}
  {Phys. Rev. Lett.}\ }\textbf {\bibinfo {volume} {113}},\ \bibinfo {pages}
  {053001} (\bibinfo {year} {2014})}\BibitemShut {NoStop}%
\bibitem [{\citenamefont {Fogarty}\ \emph {et~al.}(2013)\citenamefont
  {Fogarty}, \citenamefont {Kajari}, \citenamefont {Taketani}, \citenamefont
  {Wolf}, \citenamefont {Busch},\ and\ \citenamefont {Morigi}}]{Fogarty:2013}%
  \BibitemOpen
  \bibfield  {author} {\bibinfo {author} {\bibfnamefont {T.}~\bibnamefont
  {Fogarty}}, \bibinfo {author} {\bibfnamefont {E.}~\bibnamefont {Kajari}},
  \bibinfo {author} {\bibfnamefont {B.~G.}\ \bibnamefont {Taketani}}, \bibinfo
  {author} {\bibfnamefont {A.}~\bibnamefont {Wolf}}, \bibinfo {author}
  {\bibfnamefont {T.}~\bibnamefont {Busch}}, \ and\ \bibinfo {author}
  {\bibfnamefont {G.}~\bibnamefont {Morigi}},\ }\bibfield  {title} {\enquote
  {\bibinfo {title} {Entangling two defects via a surrounding crystal},}\
  }\href {\doibase 10.1103/PhysRevA.87.050304} {\bibfield  {journal} {\bibinfo
  {journal} {Phys. Rev. A}\ }\textbf {\bibinfo {volume} {87}},\ \bibinfo
  {pages} {050304} (\bibinfo {year} {2013})}\BibitemShut {NoStop}%
\bibitem [{\citenamefont {Brox}\ \emph
  {et~al.}(2017{\natexlab{b}})\citenamefont {Brox}, \citenamefont {Kiefer},
  \citenamefont {Bujak}, \citenamefont {Schaetz},\ and\ \citenamefont
  {Landa}}]{Brox}%
  \BibitemOpen
  \bibfield  {author} {\bibinfo {author} {\bibfnamefont {J.}~\bibnamefont
  {Brox}}, \bibinfo {author} {\bibfnamefont {P.}~\bibnamefont {Kiefer}},
  \bibinfo {author} {\bibfnamefont {M.}~\bibnamefont {Bujak}}, \bibinfo
  {author} {\bibfnamefont {T.}~\bibnamefont {Schaetz}}, \ and\ \bibinfo
  {author} {\bibfnamefont {H.}~\bibnamefont {Landa}},\ }\bibfield  {title}
  {\enquote {\bibinfo {title} {{Spectroscopy and Directed Transport of
  Topological Solitons in Crystals of Trapped Ions}},}\ }\href {\doibase
  10.1103/PhysRevLett.119.153602} {\bibfield  {journal} {\bibinfo  {journal}
  {Phys. Rev. Lett.}\ }\textbf {\bibinfo {volume} {119}},\ \bibinfo {pages}
  {153602} (\bibinfo {year} {2017}{\natexlab{b}})}\BibitemShut {NoStop}%
\bibitem [{\citenamefont {Kiethe}\ \emph {et~al.}(2021)\citenamefont {Kiethe},
  \citenamefont {Timm}, \citenamefont {Landa}, \citenamefont {Kalincev},
  \citenamefont {Morigi},\ and\ \citenamefont {Mehlst\"aubler}}]{Kiethe:2021}%
  \BibitemOpen
  \bibfield  {author} {\bibinfo {author} {\bibfnamefont {J.}~\bibnamefont
  {Kiethe}}, \bibinfo {author} {\bibfnamefont {L.}~\bibnamefont {Timm}},
  \bibinfo {author} {\bibfnamefont {H.}~\bibnamefont {Landa}}, \bibinfo
  {author} {\bibfnamefont {D.}~\bibnamefont {Kalincev}}, \bibinfo {author}
  {\bibfnamefont {G.}~\bibnamefont {Morigi}}, \ and\ \bibinfo {author}
  {\bibfnamefont {T.~E.}\ \bibnamefont {Mehlst\"aubler}},\ }\bibfield  {title}
  {\enquote {\bibinfo {title} {Finite-temperature spectrum at the
  symmetry-breaking linear to zigzag transition},}\ }\href {\doibase
  10.1103/PhysRevB.103.104106} {\bibfield  {journal} {\bibinfo  {journal}
  {Phys. Rev. B}\ }\textbf {\bibinfo {volume} {103}},\ \bibinfo {pages}
  {104106} (\bibinfo {year} {2021})}\BibitemShut {NoStop}%
\bibitem [{\citenamefont {Rotondo}\ \emph {et~al.}(2016)\citenamefont
  {Rotondo}, \citenamefont {Molinari}, \citenamefont {Ratti},\ and\
  \citenamefont {Gherardi}}]{Rotondo_2016}%
  \BibitemOpen
  \bibfield  {author} {\bibinfo {author} {\bibfnamefont {P.}~\bibnamefont
  {Rotondo}}, \bibinfo {author} {\bibfnamefont {L.~G.}\ \bibnamefont
  {Molinari}}, \bibinfo {author} {\bibfnamefont {P.}~\bibnamefont {Ratti}}, \
  and\ \bibinfo {author} {\bibfnamefont {M.}~\bibnamefont {Gherardi}},\
  }\bibfield  {title} {\enquote {\bibinfo {title} {{Devil's Staircase Phase
  Diagram of the Fractional Quantum Hall Effect in the Thin-Torus Limit}},}\
  }\href {\doibase 10.1103/PhysRevLett.116.256803} {\bibfield  {journal}
  {\bibinfo  {journal} {Phys. Rev. Lett.}\ }\textbf {\bibinfo {volume} {116}},\
  \bibinfo {pages} {256803} (\bibinfo {year} {2016})}\BibitemShut {NoStop}%
\bibitem [{\citenamefont {Sagi}\ and\ \citenamefont
  {Nussinov}(2016)}]{Nussinov:2016}%
  \BibitemOpen
  \bibfield  {author} {\bibinfo {author} {\bibfnamefont {E.}~\bibnamefont
  {Sagi}}\ and\ \bibinfo {author} {\bibfnamefont {Z.}~\bibnamefont
  {Nussinov}},\ }\bibfield  {title} {\enquote {\bibinfo {title} {Emergent
  quasicrystals in strongly correlated systems},}\ }\href {\doibase
  10.1103/PhysRevB.94.035131} {\bibfield  {journal} {\bibinfo  {journal} {Phys.
  Rev. B}\ }\textbf {\bibinfo {volume} {94}},\ \bibinfo {pages} {035131}
  (\bibinfo {year} {2016})}\BibitemShut {NoStop}%
\bibitem [{\citenamefont {Barredo}\ \emph {et~al.}(2016)\citenamefont
  {Barredo}, \citenamefont {de~Léséleuc}, \citenamefont {Lienhard},
  \citenamefont {Lahaye},\ and\ \citenamefont {Browaeys}}]{Barredo:2016}%
  \BibitemOpen
  \bibfield  {author} {\bibinfo {author} {\bibfnamefont {D.}~\bibnamefont
  {Barredo}}, \bibinfo {author} {\bibfnamefont {S.}~\bibnamefont
  {de~Léséleuc}}, \bibinfo {author} {\bibfnamefont {V.}~\bibnamefont
  {Lienhard}}, \bibinfo {author} {\bibfnamefont {T.}~\bibnamefont {Lahaye}}, \
  and\ \bibinfo {author} {\bibfnamefont {A.}~\bibnamefont {Browaeys}},\
  }\bibfield  {title} {\enquote {\bibinfo {title} {An atom-by-atom assembler of
  defect-free arbitrary two-dimensional atomic arrays},}\ }\href {\doibase
  10.1126/science.aah3778} {\bibfield  {journal} {\bibinfo  {journal}
  {Science}\ }\textbf {\bibinfo {volume} {354}},\ \bibinfo {pages} {1021--1023}
  (\bibinfo {year} {2016})}\BibitemShut {NoStop}%
\bibitem [{\citenamefont {Endres}\ \emph {et~al.}(2016)\citenamefont {Endres},
  \citenamefont {Bernien}, \citenamefont {Keesling}, \citenamefont {Levine},
  \citenamefont {Anschuetz}, \citenamefont {Krajenbrink}, \citenamefont
  {Senko}, \citenamefont {Vuletic}, \citenamefont {Greiner},\ and\
  \citenamefont {Lukin}}]{Endre:2016}%
  \BibitemOpen
  \bibfield  {author} {\bibinfo {author} {\bibfnamefont {M.}~\bibnamefont
  {Endres}}, \bibinfo {author} {\bibfnamefont {H.}~\bibnamefont {Bernien}},
  \bibinfo {author} {\bibfnamefont {A.}~\bibnamefont {Keesling}}, \bibinfo
  {author} {\bibfnamefont {H.}~\bibnamefont {Levine}}, \bibinfo {author}
  {\bibfnamefont {E.~R.}\ \bibnamefont {Anschuetz}}, \bibinfo {author}
  {\bibfnamefont {A.}~\bibnamefont {Krajenbrink}}, \bibinfo {author}
  {\bibfnamefont {C.}~\bibnamefont {Senko}}, \bibinfo {author} {\bibfnamefont
  {V.}~\bibnamefont {Vuletic}}, \bibinfo {author} {\bibfnamefont
  {M.}~\bibnamefont {Greiner}}, \ and\ \bibinfo {author} {\bibfnamefont
  {M.~D.}\ \bibnamefont {Lukin}},\ }\bibfield  {title} {\enquote {\bibinfo
  {title} {Atom-by-atom assembly of defect-free one-dimensional cold atom
  arrays},}\ }\href {\doibase 10.1126/science.aah3752} {\bibfield  {journal}
  {\bibinfo  {journal} {Science}\ }\textbf {\bibinfo {volume} {354}},\ \bibinfo
  {pages} {1024--1027} (\bibinfo {year} {2016})}\BibitemShut {NoStop}%
\bibitem [{\citenamefont {Lahaye}\ \emph {et~al.}(2009)\citenamefont {Lahaye},
  \citenamefont {Menotti}, \citenamefont {Santos}, \citenamefont {Lewenstein},\
  and\ \citenamefont {Pfau}}]{Lahaye:2009}%
  \BibitemOpen
  \bibfield  {author} {\bibinfo {author} {\bibfnamefont {T.}~\bibnamefont
  {Lahaye}}, \bibinfo {author} {\bibfnamefont {C.}~\bibnamefont {Menotti}},
  \bibinfo {author} {\bibfnamefont {L.}~\bibnamefont {Santos}}, \bibinfo
  {author} {\bibfnamefont {M.}~\bibnamefont {Lewenstein}}, \ and\ \bibinfo
  {author} {\bibfnamefont {T.}~\bibnamefont {Pfau}},\ }\bibfield  {title}
  {\enquote {\bibinfo {title} {The physics of dipolar bosonic quantum gases},}\
  }\href {\doibase 10.1088/0034-4885/72/12/126401} {\bibfield  {journal}
  {\bibinfo  {journal} {Reports on Progress in Physics}\ }\textbf {\bibinfo
  {volume} {72}},\ \bibinfo {pages} {126401} (\bibinfo {year}
  {2009})}\BibitemShut {NoStop}%
\bibitem [{\citenamefont {Baranov}(2008)}]{Baranov:2008}%
  \BibitemOpen
  \bibfield  {author} {\bibinfo {author} {\bibfnamefont {M.~A.}\ \bibnamefont
  {Baranov}},\ }\bibfield  {title} {\enquote {\bibinfo {title} {Theoretical
  progress in many-body physics with ultracold dipolar gases},}\ }\href
  {\doibase https://doi.org/10.1016/j.physrep.2008.04.007} {\bibfield
  {journal} {\bibinfo  {journal} {Physics Reports}\ }\textbf {\bibinfo {volume}
  {464}},\ \bibinfo {pages} {71--111} (\bibinfo {year} {2008})}\BibitemShut
  {NoStop}%
\bibitem [{\citenamefont {Kahan}\ and\ \citenamefont
  {Cormick}(2024)}]{Kahan:2024}%
  \BibitemOpen
  \bibfield  {author} {\bibinfo {author} {\bibfnamefont {Alan}\ \bibnamefont
  {Kahan}}\ and\ \bibinfo {author} {\bibfnamefont {Cecilia}\ \bibnamefont
  {Cormick}},\ }\bibfield  {title} {\enquote {\bibinfo {title} {Entanglement
  across the sliding-pinned transition of ion chains in optical cavities},}\
  }\href {\doibase 10.1103/PhysRevA.110.012461} {\bibfield  {journal} {\bibinfo
   {journal} {Phys. Rev. A}\ }\textbf {\bibinfo {volume} {110}},\ \bibinfo
  {pages} {012461} (\bibinfo {year} {2024})}\BibitemShut {NoStop}%
\bibitem [{\citenamefont {Laupr\^etre}\ \emph {et~al.}(2019)\citenamefont
  {Laupr\^etre}, \citenamefont {Linnet}, \citenamefont {Leroux}, \citenamefont
  {Landa}, \citenamefont {Dantan},\ and\ \citenamefont
  {Drewsen}}]{Laupretre:2019}%
  \BibitemOpen
  \bibfield  {author} {\bibinfo {author} {\bibfnamefont {T.}~\bibnamefont
  {Laupr\^etre}}, \bibinfo {author} {\bibfnamefont {R.~B.}\ \bibnamefont
  {Linnet}}, \bibinfo {author} {\bibfnamefont {I.~D.}\ \bibnamefont {Leroux}},
  \bibinfo {author} {\bibfnamefont {H.}~\bibnamefont {Landa}}, \bibinfo
  {author} {\bibfnamefont {A.}~\bibnamefont {Dantan}}, \ and\ \bibinfo {author}
  {\bibfnamefont {M.}~\bibnamefont {Drewsen}},\ }\bibfield  {title} {\enquote
  {\bibinfo {title} {Controlling the potential landscape and normal modes of
  ion coulomb crystals by a standing-wave optical potential},}\ }\href
  {\doibase 10.1103/PhysRevA.99.031401} {\bibfield  {journal} {\bibinfo
  {journal} {Phys. Rev. A}\ }\textbf {\bibinfo {volume} {99}},\ \bibinfo
  {pages} {031401} (\bibinfo {year} {2019})}\BibitemShut {NoStop}%
\bibitem [{\citenamefont {Fogarty}\ \emph {et~al.}(2015)\citenamefont
  {Fogarty}, \citenamefont {Cormick}, \citenamefont {Landa}, \citenamefont
  {Stojanovi\ifmmode~\acute{c}\else \'{c}\fi{}}, \citenamefont {Demler},\ and\
  \citenamefont {Morigi}}]{Fogarty:2015}%
  \BibitemOpen
  \bibfield  {author} {\bibinfo {author} {\bibfnamefont {T.}~\bibnamefont
  {Fogarty}}, \bibinfo {author} {\bibfnamefont {C.}~\bibnamefont {Cormick}},
  \bibinfo {author} {\bibfnamefont {H.}~\bibnamefont {Landa}}, \bibinfo
  {author} {\bibfnamefont {V.~M.}\ \bibnamefont
  {Stojanovi\ifmmode~\acute{c}\else \'{c}\fi{}}}, \bibinfo {author}
  {\bibfnamefont {E.}~\bibnamefont {Demler}}, \ and\ \bibinfo {author}
  {\bibfnamefont {G.}~\bibnamefont {Morigi}},\ }\bibfield  {title} {\enquote
  {\bibinfo {title} {{Nanofriction in Cavity Quantum Electrodynamics}},}\
  }\href {\doibase 10.1103/PhysRevLett.115.233602} {\bibfield  {journal}
  {\bibinfo  {journal} {Phys. Rev. Lett.}\ }\textbf {\bibinfo {volume} {115}},\
  \bibinfo {pages} {233602} (\bibinfo {year} {2015})}\BibitemShut {NoStop}%
\bibitem [{\citenamefont {J\"ager}\ \emph {et~al.}(2020)\citenamefont
  {J\"ager}, \citenamefont {Dell'Anna},\ and\ \citenamefont
  {Morigi}}]{Jaeger:2020}%
  \BibitemOpen
  \bibfield  {author} {\bibinfo {author} {\bibfnamefont {S.~B.}\ \bibnamefont
  {J\"ager}}, \bibinfo {author} {\bibfnamefont {L.}~\bibnamefont {Dell'Anna}},
  \ and\ \bibinfo {author} {\bibfnamefont {G.}~\bibnamefont {Morigi}},\
  }\bibfield  {title} {\enquote {\bibinfo {title} {{Edge states of the
  long-range Kitaev chain: An analytical study}},}\ }\href {\doibase
  10.1103/PhysRevB.102.035152} {\bibfield  {journal} {\bibinfo  {journal}
  {Phys. Rev. B}\ }\textbf {\bibinfo {volume} {102}},\ \bibinfo {pages}
  {035152} (\bibinfo {year} {2020})}\BibitemShut {NoStop}%
\bibitem [{\citenamefont {Abramowitz}\ and\ \citenamefont
  {Stegun}(1964)}]{AbramowitzStegun}%
  \BibitemOpen
  \bibfield  {author} {\bibinfo {author} {\bibfnamefont {M.}~\bibnamefont
  {Abramowitz}}\ and\ \bibinfo {author} {\bibfnamefont {I.~A.}\ \bibnamefont
  {Stegun}},\ }\href@noop {} {\emph {\bibinfo {title} {{Handbook of
  Mathematical Functions with Formulas, Graphs, and Mathematical Tables}}}}\
  (\bibinfo  {publisher} {Dover},\ \bibinfo {address} {New York},\ \bibinfo
  {year} {1964})\BibitemShut {NoStop}%
\bibitem [{\citenamefont {Olver}\ \emph {et~al.}(2010)\citenamefont {Olver},
  \citenamefont {Lozier}, \citenamefont {Boisvert},\ and\ \citenamefont
  {Clark}}]{Olver:2010}%
  \BibitemOpen
  \bibfield  {author} {\bibinfo {author} {\bibfnamefont {F.~W.~J.}\
  \bibnamefont {Olver}}, \bibinfo {author} {\bibfnamefont {D.~W.}\ \bibnamefont
  {Lozier}}, \bibinfo {author} {\bibfnamefont {R.~F.}\ \bibnamefont
  {Boisvert}}, \ and\ \bibinfo {author} {\bibfnamefont {C.~W.}\ \bibnamefont
  {Clark}},\ }\href@noop {} {\emph {\bibinfo {title} {The {NIST} Handbook of
  Mathematical Functions}}}\ (\bibinfo  {publisher} {Cambridge Univ. Press},\
  \bibinfo {year} {2010})\BibitemShut {NoStop}%
\end{thebibliography}%

\end{document}